\documentclass{article}
\usepackage{url}
\usepackage{amsmath}
\usepackage{mathptmx}
\usepackage{amssymb, amsmath} 
\usepackage{amscd,amsfonts}
\usepackage{theorem}
\usepackage{float}
\usepackage{amssymb, upref}
\usepackage{epsf, subfigure, verbatim}
\usepackage{latexsym}
\usepackage{mathrsfs}
\usepackage{natbib}
\usepackage{setspace}
\usepackage[T1]{fontenc}
\usepackage{times}
\usepackage[pdftex]{graphicx}

\newcommand{\ds}{\displaystyle}
\newcommand {\Rr}{\mathbb{R}}
\newcommand {\ff}{\frac{1}{2}}
\newcommand {\vrr}{V^{R}}
\newcommand {\vtt}{V^{T}}
\newcommand {\uu}{U}
\newcommand {\vv}{V}
\newcommand {\ww}{W}

\newcommand {\wrr}{W^{R}}
\newcommand {\wtt}{W^{T}}
\newcommand {\vinf}{V^{-\infty }}
\newcommand {\winf}{W^{-\infty }}

\newcommand {\ir}{i_{R}}

\newcommand{\ddiv}{\textrm{div}\,}
\newcommand{\tr}{\tilde{R}}

\newcommand{\cov}{\text{cov}}
\newcommand{\var}{\text{var}}

\newcommand{\V}{{V}}
\newcommand{\W}{{W}}
\newcommand{\subV}{{V}}
\newcommand{\subW}{{W}}
\newcommand{\X}{{X}}
\newcommand{\thrV}{{\V^T}}
\newcommand{\thrW}{{\W^T}}
\newcommand{\thrX}{{\X^T}}
\newcommand{\CCG}{{R}}

\newtheorem{proposition}{Proposition}
\newtheorem{remark}{Remark}

\date{\today}

\begin{document}

\title{Finite volume and asymptotic methods for stochastic neuron models with correlated inputs}

\author{Robert Rosenbaum, Jianfu Ma, Fabien Marpeau, Aditya Barua and Kre\v simir Josi\'c} 

\maketitle

\begin{abstract}
We consider a pair of stochastic integrate and fire neurons receiving correlated stochastic inputs.
The evolution of this system can be described by the corresponding Fokker-Planck equation with 
non-trivial boundary conditions resulting from 
 the refractory period and firing threshold.  We
propose a finite volume method that is orders of magnitude faster than the Monte Carlo methods
traditionally used to model such systems. 
The resulting numerical approximations are proved to be accurate, nonnegative and integrate to 1.   
We also approximate the transient evolution of the system using an Ornstein--Uhlenbeck process,
and use the result to examine the  properties of the joint output of cell pairs. The results suggests that the joint output of 
a cell pair is most sensitive to changes in input variance, and less sensitive to changes in
 input mean and correlation.
\end{abstract}



\section{Introduction}
\label{intro} The integrate and fire (IF) model is used widely in mathematical biology~\citep{burkitt06i,keener}. It is
simple, yet versatile, and provides a good approximation of the
response of an excitable cell in a variety of situations. A
stochastic version of the IF model can  describe the behavior of large populations of
cells through the evolution of the corresponding
probability density~\citep{Knight72,Nykamp00,Deco08}.  It can also be used
to study the response of a single cell subject to a large number of small,
statistically independent inputs~\citep{lindner01,Renart03}.

Collections of excitable cells frequently do not behave independently.
The joint response of populations of electrically active cells is
of interest in a number of areas in biology: Pancreatic
$\beta$-cells have to synchronize their response to secrete
insulin~\citep{meda,sherman91}, and the coordinated activity of
cardiac cells is essential for their function~\citep{keener}.  Our
study is motivated primarily by cells in neural populations. Such
cells typically fire action potentials in response to synaptic
inputs from other cells.  The standard stochastic IF model can capture
the response of a cell when such inputs are independent~\citep{Renart03}. However
dependencies between these inputs cannot always be ruled out.  Such dependencies can affect the output statistics of a neuronal population, and significantly impact the amount
of information carried in the population response~\citep{salinas00,sompolinsky01}. Even weak
correlations between individual cells can significantly impact the
ensemble activity of a population~\citep{shadlen98,rosenbaum10}.
Here we describe a model that can be used to examine the behavior of two cell
populations (or cell pairs) receiving correlated inputs.

We first develop a
Fokker-Planck equation that describes the evolution of the
probability density for a pair of cells receiving correlated inputs.   The response of cell pairs
receiving correlated inputs has been studied previously using linear response
theory~\citep{ostojic,rocha07}, and numerical simulations~\citep{galan07} in related models.
However, the boundary conditions in the presence of a
refractory period are nontrivial, and can impact the behavior of the system.  We therefore
present the model in some detail.

Next we describe a finite volume method that can be used to 
study solve the Fokker-Planck equation numerically for the probability density.  Previously, we proposed a
fast and accurate finite volume method for modeling a general IF
neuron driven by a stochastic 
input~\citep{faj_neuro}.  This method was significantly faster than
Monte Carlo (MC) simulations and we proved several stability properties of the algorithm. Here we extend this method to two neurons with correlated inputs.
While the dynamics of interacting populations has been examined previously~\citep{Nykamp00,harrison05}, we are not aware of a numerical treatment of the Fokker--Planck
equation corresponding to stochastic IF
neurons driven by correlated noise.  

Finally, we develop a simple analytical approximation in terms of a
related Ornstein-Uhlenbeck process that captures the
response of a cell pair  to study the behavior
of cells, or cell populations receiving correlated inputs.
This approach provides an alternative to the linear response
techniques commonly in use~\citep{lindner01,ostojic}.

We use this approximation to examine the response of a single cell and a cell pair to
changes in the input parameters.   The variance (noisiness), mean and synchrony
between the inputs are separate channels along which information
can be communicated to postsynaptic cells.
We find that the spiking statistics of a single cell and
the cell pair are most sensitive to changes in the variance of the input.    This
suggests that the joint response of a cell population most accurately 
tracks input noise intensity.  

\section{Model Description}

\label{description}

A single IF neuron with stochastic input is described by
the Langevin equation:
\begin{equation} \label{eq:Lan}
\frac{dV}{dt} =f(V)+\sqrt{2D} \xi(t), \qquad \qquad V \in (-\infty, \vtt).
\end{equation}
Here $f$ defines the deterministic (drift) behavior, and  $\xi(t)$
 a Gaussian stochastic processes with
$\langle \xi(t)  \rangle =0$ and $\langle \xi(t)\xi(t^{\prime}) \rangle =\delta(t-t^{\prime})$.
When the voltage reaches a threshold,  $\vtt$, a spike is fired, and $V$ is
instantaneously reset to $\vrr < \vtt$.
A spike may be followed by an absolute refractory period $\tau$,  during which
a neuron is insensitive
to inputs, and   $V$ is held fixed at $\vrr$.

\begin{figure*}
\begin{center}
\includegraphics[width=0.65\textwidth,height=0.4\textwidth]{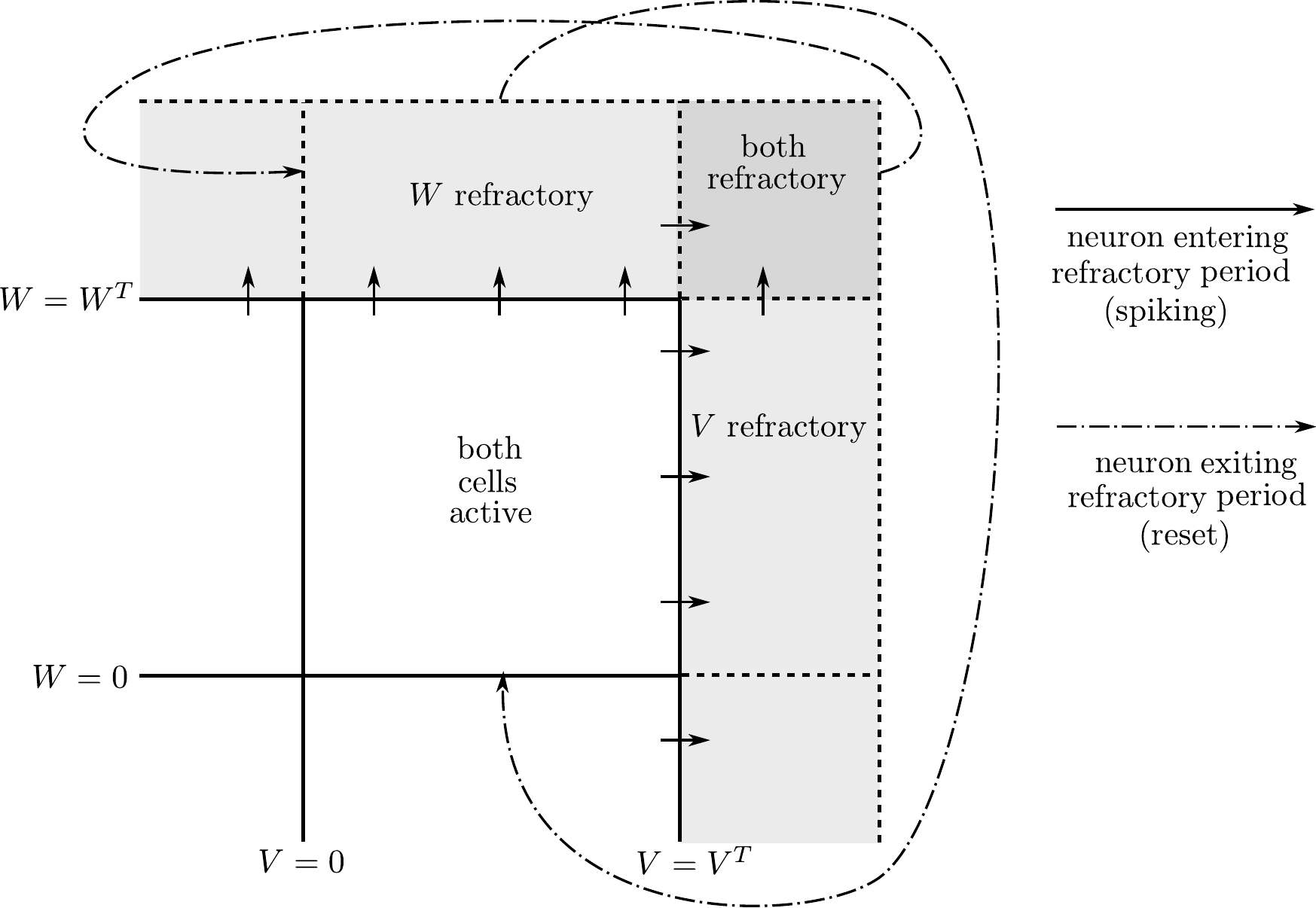}
\end{center}
\caption{ (Left) Domain of simulation. (Right) Circulation of probability mass through populations $P$, $R_{\vv }$, $R_{\ww }$ and $R$. }
\label{fig:domain}       
\end{figure*}

This model can also be understood
as a the diffusive limit of a population of cells receiving independent inputs~\citep{Omurtag00}.
To model a pair of cells receiving correlated inputs, we assume that their membrane voltages $V$ and $W$  obey a
pair of Langevin equations:
\begin{equation}  \label{E:model1}
\begin{split}
         \dot V &= f \left(V, W \right) + I_V(t)  \, ; \quad I_V(t)  =  \mu_V+  \sqrt{2D}(\sqrt{1-c} \xi_V(t) + \sqrt{c} \xi_c(t))    \\
        \dot W &= g \left(W, V \right) + I_W(t) \,  ; \quad I_W(t)  = \mu_W+  \sqrt{2D}(\sqrt{1-c} \xi_W(t) + \sqrt{c} \xi_c(t)).
\end{split}
\end{equation}
The inputs, $I_j(t)$, received by the cells are comprised of statistically {\it independent} stochastic processes $\xi_V(t)$ and $\xi_W(t)$,  and a common input $\xi_c(t)$.
The $\xi_i$ are again assumed to be Gaussian with
$\langle \xi_i(t)  \rangle =0$ and $\langle \xi_i(t)\xi_j(t^{\prime}) \rangle =\delta(t-t^{\prime})\delta_{i,j}$. The constant $c$, is the Pearson correlation coefficient
between the inputs and lies between 0 and 1. For instance, for two  leaky integrate-and-fire (LIF) neurons with common input, but no direct coupling, $f(V,W)=-g_V (V - V_{\text{rest}})$ and $g(V,W) = -g_W (W - W_{\text{rest}})$.  Each cell spikes when the voltage $V$ crosses the threshold, $\vtt$ and $\wtt$ respectively.
After each spike the voltage is reset to $\vrr < \vtt$ ($\wrr < \wtt$ for cell 2) , and is pinned to this value for the duration of
the refractory period,  $\tau _{\vv}$ ($\tau _{\ww}$ for the second cell, see Fig.~\ref{fig:domain}).  For simplicity
we will refer to the two as neuron $V$ and $W$, although this can be understood as ``populations $V$ and $W$''~\citep{harrison05}.
The joint probability density of the two voltages evolves on the domain $\Omega = (\vinf,\vtt )\times (\winf ,\wtt )$. In theoretical studies it is frequently assumed that  $\vinf = \winf = -\infty$.
However, since we will be interested in numerical simulations, we assume that these quantities are large and negative.

With  $\uu =(V,W)$ and $F=(f,g)$ the Fokker-Planck equation corresponding to Eq.~\eqref{E:model1} takes the form
\begin{equation}
\partial _{t}P(t,\uu )+\ddiv \big( F(\uu )P(t,\uu ) - D M\nabla P(t,\uu )\big) =0, \qquad M=\left( \begin{array}{cc}
1 & c\\
c & 1
\end{array}
\right), \label{eq:fpgeneral}
\end{equation}
for $\vv  \in (\vinf,\vtt ) \backslash \vrr $, $\ww \in (\winf ,\wtt ) \backslash \wrr $.  Here  $D$ is the
diffusion coefficient and $M$ is the correlation matrix.
This equation is coupled with reflecting boundary conditions at $\vv =\vinf$ or $\ww =\winf$,  and absorbing boundary conditions at both thresholds:
\begin{equation}
\begin{split}
\big( f(\uu ) - D\partial _{\vv }-cD\partial _{\ww } \big) P(t,\uu )\big|_{\vv =\vinf} & = 0, \qquad
\big( g(\uu ) - D\partial _{\ww }-cD\partial _{\vv } \big) P(t,\uu )\big|_{\ww =\winf}  = 0,\\
P(t,\uu )|_{\vv =\vtt }=P(t,\uu )\big|_{\ww =\wtt } & = 0  .
\label{eq:absorbing}
\end{split}
\end{equation}

The presence of the refractory behavior in the IF model introduces additional complexity.  If either neuron enters its refractory state, the corresponding voltage is fixed at the reset value, and the entire system  effectively evolves according to a one-dimensional Fokker-Planck equation. During this time, it is possible that the second  cell also crosses the threshold, fires and enters the refractory state.  In this case the voltages are fixed at $(\vrr, \wrr)$ and both neurons are insensitive to inputs until one of them exits the refractory state.

To capture the behavior of neurons in the refractory period, we
model the evolution of densities using three separate, communicating
sub-populations, in addition to $P(t,
\uu)$~\citep{Sirovich:2008p1751,Ly:2009p1340}:
\begin{itemize}
\item $R_{\vv }(t,r,\vv )$,
the probability density of the fraction of the population in which only neuron $W$ is in the refractory state,
\item $R_{\ww }(t,s,\ww )$ the corresponding density in which only neuron $V$ is in the refractory state, and
\item $R(t,s,r)$, the  density corresponding to both neurons in the refractory state.
\end{itemize}
For all densities, $t$ refers to the time since the beginning of the simulation, while $r$ and $s$
refer to relative times measured from the beginning of the refractory period for neuron $V$ and $W$, respectively.
Therefore, $R_{\vv}(t,r,\vv_0)\Delta \vv \Delta r$ is the fraction of the population for which neuron
$W$ has been in the refractory period between $r$ and $r+\Delta r$ units of time, and the
voltage of neuron $V$ is between $\vv_0$ and $\vv_0 + \Delta \vv$.
The quantity
 $R(t, s,r)\Delta s \Delta r$
is the fraction of the population in which neurons $V$ and $W$ have
been in refractory periods for times in the intervals  $[s, s+\Delta
s]$ and $[r, r+\Delta r]$ respectively. The use of variables $s$ and
$r$ is closely related to age-structured population dynamics
models~\citep{iannelli,webb}. Indeed, $s$ and $r$ denote the
``ages'' of the refractory states for neurons 1 and 2 respectively.
Fig.~\ref{fig:domain} summarizes the circulation of probability mass
between the different populations involved.

 Since
the entire population is described by these densities we have for any time $t$
\begin{multline}
\int _{\vinf }^{\vtt }\int _{\winf }^{\wtt } P(t,V,W)\, dW\, dV + \int _{0}^{\tau _{\ww}}\int _{\winf }^{\wtt }R_{\ww }(t,s,W)\, dW\, ds \\
+ \int _{0}^{\tau _{\vv}}\int _{\vinf }^{\vtt }R_{\vv }(t,s,V)\, dV\, ds + \int _{0}^{\tau _{\vv}}\int _{0}^{\tau _{\ww}} R(t,s,r)\, ds \, dr = 1. \label{eq:conservation}
\end{multline}

We next describe the evolution of the main population and the three refractory populations and how they are coupled to
each other through boundary terms.  The refractory populations
evolve according to one-dimensional Fokker-Planck equations,
\begin{equation}
\begin{split}
(\partial _{t}&+\partial _{r})R_{\vv }(t,r,\vv )+\ddiv \big( f(\uu )R_{\vv }(t,r,\vv ) - D\partial _{\vv }R_{\vv }(t,r,\vv )\big) =0  ,r\in (0,\tau _{\ww }),  \\
(\partial _{t}&+\partial _{s})R_{\ww }(t,s,\ww )+\ddiv \big( g(\uu
)R_{\ww }(t,s,\ww ) - D\partial _{\ww }R_{\ww }(t,s,\ww )\big) =0
,s\in (0,\tau _{\vv }), \label{eq:bc2d2}
\end{split}
\end{equation}
\begin{equation}
(\partial _{t}+\partial _{s}+\partial _{r})R(t,s,r)=0. \label{eq:bc2d4}
\end{equation}

In addition, mass from the main population $P(t, \uu )$ is injected into
populations  $R_{\vv }(t,r,\vv )$ and $R_{\ww }(t,s,\ww )$ at $r = 0$ and $s = 0$
as neuron $V$ and $W$ cross threshold respectively.  These source terms are described by
\begin{equation}
R_{\vv }(t,0,\vv )=-D\partial _{\ww }P(t,\vv ,\wtt ),\ R_{\ww }(t,0,\ww )=-D\partial _{\vv }P(t,\vtt ,\ww ) \ .\label{eq:bc2d1}
\end{equation}
 While either neuron is in the refractory state, the other neuron can enter its own refractory state as well, providing boundary conditions to equation (\ref{eq:bc2d4}) for inward characteristics:
\begin{equation}
R(t,s,0)=-D\partial _{\ww }R_{\ww }(t,s,\wtt ),\ R(t,0,r)=-D\partial _{\vv }R_{\vv }(t,r,\vtt )\ .\label{eq:bc2d1prime}
\end{equation}
Next, while both neurons are in the refractory state, neuron $V$ or
$W$ may exit the refractory state while the other neuron remains in
the refractory state.  Using  $[\xi ]\big|_{z = Z}:= \lim _{z\to
Z^{+}}\xi (z) - \lim _{z\to Z^{-}}\xi (z)$ to denote the jump across
point $Z\in \Rr $, we can express the contribution of population
$R(t,s,r)$ to the following source
terms~\citep{Melnikov:1993p1752,lindner01}
\begin{equation}
\big[ D\partial _{\vv } R_{\vv }(t,s,.) \big]\bigg|_{\vv =\vrr } = R(t,\tau _{\vv },s),\qquad \big[ D\partial _{\ww } R_{\ww }(t,r,.) \big]\bigg|_{\ww=\wrr } = R(t,r,\tau _{\ww }). \label{eq:bc2d5}
\end{equation}
As either neuron exits the refractory period, it re-enters the main population modeled
by the density $P(t, \uu)$.  This is captured by adding the source terms:
\begin{equation}
\big[ D\partial _{\vv } P(t,.,\vv ) \big] \bigg|_{\vv =\vrr } = R_{\ww }(t,\tau _{\vv },\ww ),\qquad \big[ D \partial _{\ww } P(t,\vv ,.)\big]\bigg|_{\ww =\wrr } = R_{\vv }(t,\tau _{\ww },\vv ) \ .\label{eq:bc2d6}
\end{equation}
The densities are also continuous across the reset potentials, so that
\begin{equation}
\big[ P(t,.,\ww ) \big] \bigg|_{\vv =\vrr }=\big[ P(t,\ww ,.) \big]\bigg| _{\ww =\wrr }=\big[ R_{\vv }(t,s,.) \big] \bigg|_{\vv =\vrr }=\big[ R_{\ww }(t,s,.) \big]\bigg|_{\ww =\wrr }=0\ .
\label{eq:bc2d7}
\end{equation}
Finally, reflecting and absorbing boundary conditions are imposed on
Eq.~(\ref{eq:bc2d2}), and (\ref{eq:bc2d4}) by requiring:
\begin{equation}
\big( f(\uu ) - D\partial _{\vv } \big) R_{\vv }(t,s,. )_{|\vv =\vinf}=\big( g(\uu ) - D\partial _{\ww }\big) R_{\ww }(t,s,.)_{|\ww =\winf} = 0\ ,
\label{eq:bc2d8}
\end{equation}
\begin{equation}
R_{\vv }(t,s,\vtt )=R_{\ww }(t,s,\wtt )=0\ .
\label{eq:bc2d9}
\end{equation}
Notice that when there is no refractory period ($\tau _{\vv }=\tau _{\ww}=0$), (\ref{eq:bc2d1})--(\ref{eq:bc2d6}) reduce to the single boundary condition
\begin{equation}
\big[ D\partial _{\vv } P(t,.,\ww ) \big] _{|\vv =\vrr } = -D\partial _{\vv }P(t,\vtt ,\vv )\ ,\  \big[ D \partial _{\ww } P(t,\vv ,.)\big]_{|\ww =\wrr } = -D\partial _{\ww }P(t,\vv ,\wtt ) \ .\nonumber 
\end{equation}

\section{Description of the Numerical Methods}

The numerical methods used in simulating the solutions of the model described
in the previous section are not completely standard.  The anisotropy of the diffusion operator
coupled with the absorbing boundary condition presents numerical challenges different
from those encountered, for instance, when modeling phase oscillators~\citep{galan07}.
We therefore give a brief
description of our approach here.

\subsection{Finite Volume Method}

Three requirements in the numerical discretization of Eqs.~(\ref{eq:fpgeneral})--(\ref{eq:bc2d9}) are obtaining numerical
probability densities which are accurate, nonnegative and integrate
to 1.  We dealt with similar difficulties in \citep{faj_neuro}, and
use an extension of that approach here.  To obtain numerical
densities which integrate to $1$, we use conservative numerical
schemes which ensure that the mass lost by a mesh-element is
transmitted exactly to its neighboring elements.  This ensures
preservation of mass of the initial densities. Secondly, the drift
operator is well known to be an obstacle when trying to combine
accuracy, non-negativity, and stability of the numerical densities.
Therefore, we use an operator splitting method (described below)
that enables us to discretize the drift and diffusion operators
separately. The discretization of the drift term is carried out by
an upwind scheme whose accuracy is improved using flux limiters.

However, the two-dimensional nature of the problem induces further difficulties:
\begin{enumerate}
\item Extra numerical diffusion is generated in the cross directions from the drift operator, leading to a loss of accuracy.
\item Due to the correlation coefficient $c$, the diffusion matrix $DM$ is anisotropic. The discretization of the cross derivatives $\partial _{\vv \ww}^{2}P$ commonly involves the inversion of matrices that are not unconditionally strongly diagonally dominant, which makes it difficult to obtain nonnegative numerical densities.
\item Due to the refractory periods, the multiple boundary conditions (\ref{eq:bc2d2})--(\ref{eq:bc2d9})
drastically increase the algorithmic complexity, compared with the
one-dimensional model in \citep{faj_neuro}. In particular, the
densities $R_{\vv }$ and $R_{\ww}$ have to be obtained by
discretizing Eq.~(\ref{eq:bc2d2}) at each time step and age step.
Moreover, the code has to gather all the phenomena and assemble the
circulation between the four populations in an efficient manner.
\end{enumerate}
The technical of our approach are relegated to
Appendix~\ref{numerical}, and what follows is an outline. First, the
time interval $\Rr _{+}$ on which the solution will be approximated
is partitioned into sub-intervals $(t_n,t_{n+1})$. We will denote
the numerically obtained approximation of the population $P$ at time
$t_{n}$ by $P^{n}$. As in \citep{faj_neuro}, $P^{n+1}$ is obtained
from $P^n$ by splitting equations
(\ref{eq:fpgeneral})--(\ref{eq:absorbing}) into
\begin{equation}
\partial _{t}P+ \ddiv (fP) = 0,\ P\big|_{\vv =\vinf }=P\big|_{\ww =\winf }=0,\label{eq:transport}
\end{equation}
\begin{equation}
\begin{array}{c}
\partial _{t}P - \ddiv (DM\nabla P) = 0,\\
\big( D\partial _{\vv }+CD\partial _{\vv \ww } \big) P\big|_{\vv =\vinf}=\big( D\partial _{\ww }+CD\partial _{\vv \ww } \big) P\big|_{\ww =\winf} =P\big|_{\vv =\vinf }=P\big|_{\ww =\winf }= 0.
\end{array}
\label{eq:diffreac}
\end{equation}
The numerical solution is updated at time $t_{n+1}$ using
\begin{equation}
P^{n+1}=\mathcal{S}_{2}\big( \mathcal{S}_{1}(P^{n})\big) ,\label{eq:splitting}
\end{equation}
where $\mathcal{S}_{1}$ and $\mathcal{S}_{2}$ are approximation schemes for Eq.~(\ref{eq:transport}) and (\ref{eq:diffreac}) respectively, along with split interior conditions specified later. This technique allows us to develop specific numerical schemes which are adapted to each differential operator in Eq.~(\ref{eq:fpgeneral}).

The numerical scheme $\mathcal{S}_{1}$ is nonlinear explicit. A
compromise between accuracy and stability is obtained by adding flux
limiters to the upwind scheme. As discussed in the Appendix, we use
the numerical scheme introduced in~\citep{faj_neuro} in each
direction, $V$ and $W$. The time step is restricted by a
Courant-Friedrichs-Lewy ({\sc{CFL}}) condition~\citep{cfl,god},
which provides stability and positivity preservation of the scheme
by ensuring that the drift term does not shift the numerical
solution by more than one mesh element per time step~\citep{god}.

The scheme $\mathcal{S}_2$ is linear implicit. Using centered approximation of the derivatives $\partial _{\vv \vv }P$, $\partial _{\ww \ww }P$ and a semi-center discretization of the cross derivative (see \ref{subsection:diffusion}), the numerical solution is obtained by inverting a matrix that is strongly diagonally dominant as long as the mesh-size does not change too sharply between two elements. The scheme will remain stable and positivity preserving with any time step. The inversion of the linear system is carried out by an LU pre--conditioned gradient procedure.

The one dimensional Fokker--Planck equations (\ref{eq:bc2d2}) are
discretized by using the one-dimensional scheme from
\citep{faj_neuro} (See~\ref{subsection:1dfp}). The main difference
here is the presence of the age variables, $s$ and $r$. Since the
age evolves simultaneously with time, we just solve
Eq.~(\ref{eq:bc2d2}) at each time step, regardless of age, and shift
the age variable by one time step. Finally, Eq.~(\ref{eq:bc2d4}) is
solved exactly. The other boundary conditions given in
Eqs.~(\ref{eq:bc2d1})--(\ref{eq:bc2d9}) are discretized as in one
space dimension. The convergence criterion is satisfied as the
residual of the numerical scheme decreases to a pre-defined value,
$10^{-6}$ in our study.



\subsection{Computing spike train statistics}\label{S:spikestats}

The statistics of the number of threshold crossings of an IF model are of special interest.  Using neuroscience terminology, we will refer to each threshold crossing  as a \emph{spike} and the sequence of threshold crossings as a \emph{spike train}.
Let the stochastic set functions $N_V(s,t)$ and $N_W(s,t)$ denote the number spikes during the time interval $[s,t]$ in cell $V$ and $W$, respectively.   The \emph{instantaneous firing} rate of cell $X=V,W$ is defined
as the instantaneous rate at which the corresponding population of cells spikes
at time $t$.  In terms of the spiking probability of a single cell, this can be written as
$$
\nu_X(t)=\lim_{\Delta t\to 0}\frac{1}{\Delta t}\Pr\left(N_X(t,t+\Delta t)>0\right).
$$
The \emph{conditional firing rate}, $\nu_{V|W}(\tau,t)$, is defined as the firing rate of cell $V$  at time $t+\tau$ given that  $W$ has spiked at time $t$,
$$
\nu_{V|W}(\tau,t)=\lim_{\Delta t\to 0}\frac{1}{\Delta t} \Pr\left(N_V(t+\tau+\Delta t,t+\tau)>0\,\big |\, N_W(t,t+\Delta t)>0\right)
$$
and similarly for $\nu_{W|V}(\tau,t)$.
The conditional firing rate can be normalized by the rates to obtain the spike train \emph{cross-covariance function}
\begin{equation}\label{E:CCGdef}
C_{\vv\ww}(\tau,t)=\nu_W(t)\left(\nu_{V|W}(\tau,t)-\nu_V(t+\tau)\right),
\end{equation}
which is a common measure of correlation between the activity of two neurons over time.
In the study of neural coding, it is often useful to know the propensity of one cell to spike during some time interval after another cell has spiked.  For this purpose, we define the conditional mean rate,
\begin{equation}\label{E:En}
S_{V|W}(a,b,t)
=\frac{1}{(b-a)}\int_a^b \nu_{V|W}(t,\tau)d\tau
\end{equation}
When the distribution of membrane potentials is in steady state, the spike trains are stationary and we can
drop the explicit dependence on $t$ to write $\nu_X$, $\nu_{V|W}(\tau)$, $C_{VW}(\tau)$, and $S_{V|W}(a,b)$ without ambiguity.

Since action potentials are not explicitly modeled in the Fokker-Planck formalism described above, the spiking statistics must be calculated using properties of the probability density near threshold.
The instantaneous firing rate of cell $\vv$ can be obtained from the solution of the Fokker-Planck equation by taking the marginal flux over threshold,
\begin{eqnarray} \label{MglFlux}
\nu_\vv(t)=-D \,P_\vv(t,\vv)\big|_{\vv=\vv^T}
\end{eqnarray}
where $P_\vv(t,\vv)=\int_{\winf}^{\ww^T} P(t,\vv,\ww)d\ww$ is the
marginal density of $\vv$.  Thus, up to terms of $O((\Delta t)^2)$,  the quantity $\nu_V(t) \Delta t$ is the probability mass that crossed threshold during the interval $\Delta t$.  Equivalently, it equals the probability that a cell fires during this interval.   The instantaneous firing rate of $\ww$
is defined analogously.

The conditional firing rates are obtained by first calculating the conditional flux immediately after a spike in cell $W$ at time $t$,
\begin{eqnarray} \label{CdlFlux}
J_{\text{cond}}(t,V):= -D\, \partial_\ww
P(t,\vv,\ww)\big|_{\ww=\wtt}.
\end{eqnarray}
This conditional flux is then normalized to give the conditional density,
$J_{\text{cond}}(t,V)\to J_{\text{cond}}(t,V)\left/
\int_{\vinf}^{\vtt}J_{\text{cond}}(t,x)dx\right.$ and used as an
initial condition for the 1-dimensional Fokker-Planck equation,
\begin{eqnarray} \label{E:F1d}
\partial_{\tau} P_{1}(V,\tau)=-\partial_\vv (f(V)-D\,\partial_{V} P_{1}(V,\tau)).
\end{eqnarray}
As the solution of this equation evolves, the conditional firing rate of $\vv$ is given by $\nu_{V|W}(\tau,t)=-D\,\partial_{V}P_{1} (V, \tau)$.
The conditional firing rate can then be normalized, \emph{cf.} Eq.~\eqref{E:CCGdef}, to get the cross-covariance function, or integrated, \emph{cf.} Eq.~\eqref{E:En}, to get the conditional expected mean rate.
We experienced convergence problems with the derivative of the finite volume solutions at the upper corner, $(\vv^T, \ww^T)$, of the spatial domain.
Due to the convergence issues discussed in Appendix \ref{A:FVspikestats}, the finite volume approximation to the conditional firing rate does not converge when $\tau$ is very small.

\section{Validation of the numerical solution}

As the  finite volume
numerical scheme  we developed is novel, we first  compare its output
to that obtained using Monte Carlo (MC)  simulations (See Appendix~\ref{MC}).   We
consider both stationary and non-stationary inputs.

\begin{figure}[!ht]
\begin{center}
\includegraphics[scale=0.7]{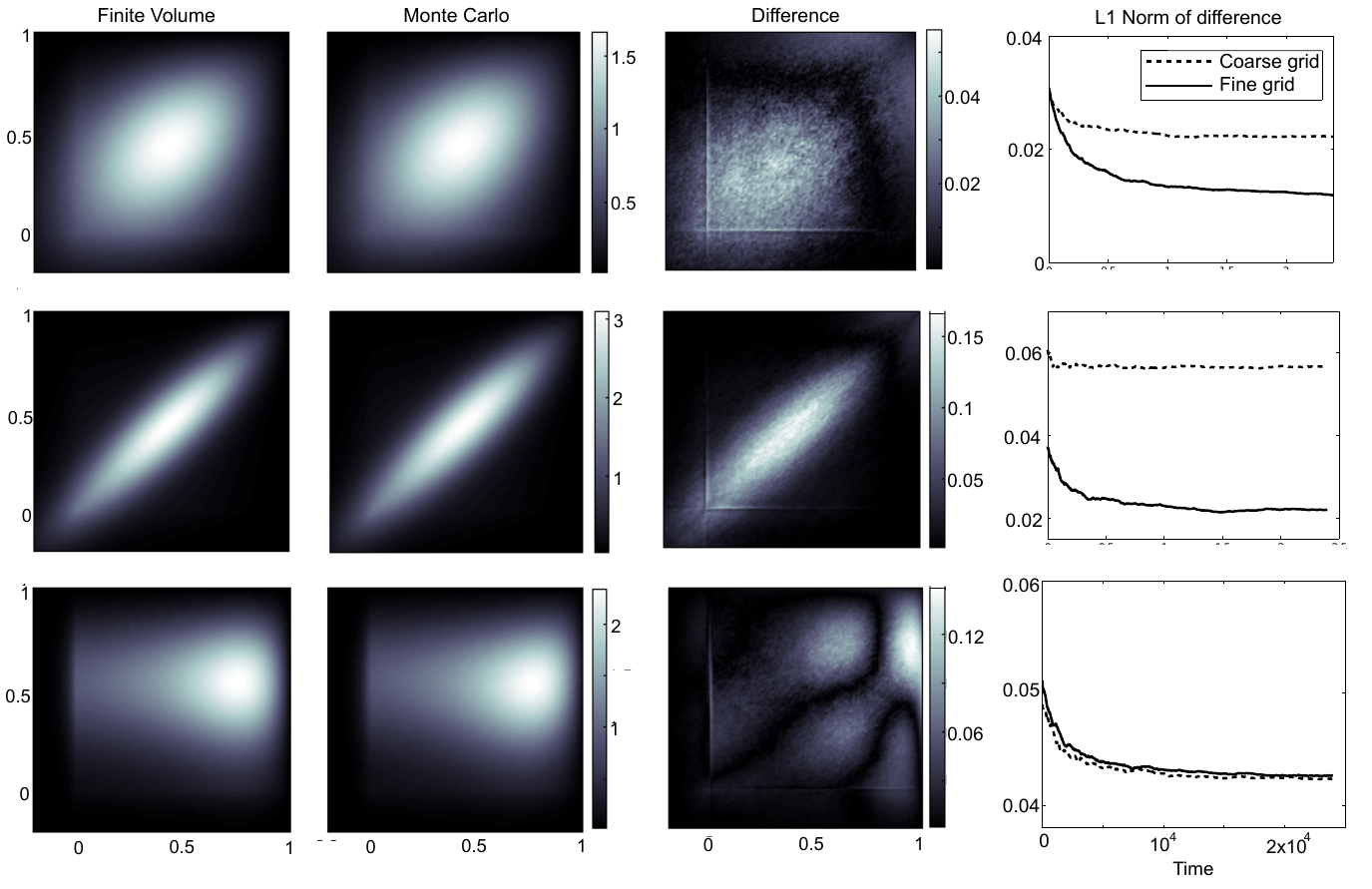}
\end{center}
\caption{\emph{Stationary probability densities for a pair of LIF
neurons}:  Results from finite volume simulations (first), MC
simulations (second), the difference between the two approximations
(third column).  To test convergence of the finite volume method, we
used a coarse ($100 \times 100$ elements in the unit square),
 and a fine grid ($200 \times 200$ elements). The fourth column
 shows the $L^1$ norm of
the difference between the equilibrium distributions obtained using
the finite volume and MC simulations. Parameters: for the top panel,
$\mu_{W}=\mu_{V}=0.5$, $D=0.05$, $c=0.5$, $\tau=0.5$; for the middle
panel, $\mu_{W}=\mu_{V}=0.5$, $D=0.05$, $c=0.9$, $\tau=0.5$; for the
bottom panel, $\mu_{V}=1.2, \mu_{W}=0.6$, $D=0.05$, $c=0.3$,
$\tau=0.2$.  The first three columns were obtained using a coarse ($100\times 100$) grid.
}
\label{FigTimeIndep}
\end{figure}

As an example we choose the case of two LIF  neurons which
corresponds to setting $f(V, W)=-V+\mu_{V}$, and $g(V,W)=-W+\mu_{W}$
in Eq. ($\ref{eq:fpgeneral}$)~\citep{burkitt06i}.   When $\mu_{V}<
V^{T}$ and $\mu_{W}< W^{T}$ the cells are in the fluctuation
dominated regime, and firing is due to large excursions of membrane
voltages from the mean. As shown in top and middle panels of
Fig.~\ref{FigTimeIndep}, the finite volume method provides an
excellent approximation of the stationary distribution when the
input to the two cells is constant in time.
 As the correlation between the inputs to the two cells,
 measured by $c$, increases, the membrane potentials become more correlated, and
 their joint  probability
density is stretched along the diagonal.

When $\mu_{V}> V^{T}$ or $\mu_{W}> W^{T}$ it is the DC component of
the input current that drives the cells over threshold.  This situation is
somewhat more challenging to simulate, since much of the mass of the
invariant distribution lies close to the threshold.
The gradient of the solution close to the boundary becomes large.  Together with
the Dirichlet boundary conditions, this causes larger errors in the numerical approximation
close to the boundary. The
bottom panels of Fig.~\ref{FigTimeIndep} show that the finite volume
method still performs well in  this situation.

\begin{figure}[!ht]
\begin{center}
\includegraphics[scale=1]{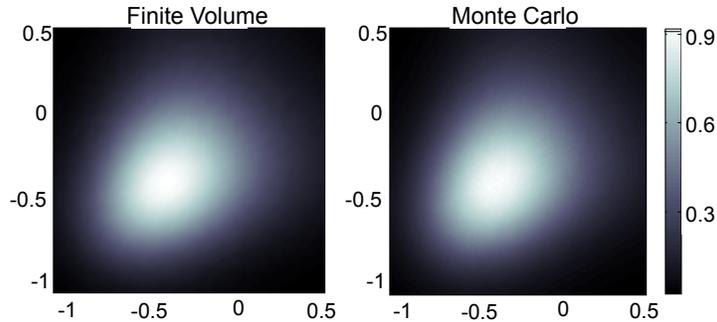}
\end{center}
\caption{\emph{Stationary probability densities for a pair of QIF
neurons}:  finite volume method (left), MC simulations (right) when
$\mu_{V}=\mu_{W}=-0.1$, $D=0.1$ and $c=0.3$.} \label{FigQIF}
\end{figure}

We can change the drift term in Eq.~\eqref{E:model1} to simulate a
different integrate--and--fire model. In particular, the quadratic
integrate and fire (QIF) model is obtained by setting $f(V,
W)=V^2+\mu_{V}, g(V,W)=W^2+\mu_{W}$~\citep{ermentrout86,brunel03}.
Fig.~\ref{FigQIF} demonstrates that the finite volume numerical
scheme performs well in computing the invariant distribution for
this model.

The finite volume scheme was designed to compute the evolution of
the joint probability density of the two sub-threshold voltages in
time. Stationary distributions were presented here for ease of
visualization. A comparison of time dependent solutions obtained
using finite volume and MC methods is  available online at
\url{http://www.math.uh.edu/~josic/myweb/research/papers/FV/}.
The animation shows the time dependent density from $t=100$ to $t=110$ for a pair of LIFs with $\mu=|\sin(t)|$, $c=|\sin(t)|/2$, and $D=0.1 |\sin(t)|$.

\section{Gaussian approximation}

The LIF model is ubiquitous in stochastic modeling of excitable
systems primarily due to its mathematical tractability.  Closed form
expressions have been obtained for the stationary density, spiking
statistics, and linear response properties of the one neuron
model~\citep{lindner01}.  However, exact closed form expressions are
not known for the two neuron model with $c\ne 0$ discussed in
Section~\ref{description}.  The numerical methods we describe here
offer a way of exploring the behavior of the LIF model in the
absence of analytic solutions. However, even with fast numerical
methods, exploring large regions of parameter space may not be
possible.  Approximate analytic solution, are therefore frequently
necessary to gain a deeper understanding of the model.

 Much recent work has focused on deriving such approximations using perturbative methods.  Linear response theory was used to study the dependencies in the output of a call pair receiving correlated input~\citep{lindner01,ostojic,rocha07,SheaBrown:2008p1754}.  These solutions involve integrals that must be evaluated numerically.  Simpler approximations can be obtained by ignoring the threshold and reset boundary conditions
 when the neurons are in the fluctuation dominated regime, and firing rates are low.  Neurons in the cortex may reside in this regime
under typical conditions~\citep{Ringach:2007p1217}. Previous approximations obtained in this regime 
required smoothness assumptions on the trajectories of the membrane potentials~\citep{burak09,tchumatchenko10}.
Since solutions to Eq.~\eqref{E:model1} are nowhere differentiable when
$D>0$, a different approach must be used for the LIF driven by white
noise inputs. We next describe such an approximation.  (We note that a similar
approach has been used to examine the response of integrate-and-fire
neurons driven by filtered Gaussian noise~\citep{badel10}.)

When firing rates are small, the boundary conditions have a
small impact on the solution of Eq.~\eqref{eq:fpgeneral} and an
approximate solution can be obtained by solving the free boundary
problem ($\vtt, \wtt \rightarrow \infty$).  Since firing is rare,
the amount of time spent in the refractory states is negligible and
the refractory period can be ignored ($\tau=0$). Under this
approximation, the stochastic process $(V(t),W(t))$ is an
Ornstein-Uhlenbeck process in $\Rr^2$~\citep{gardiner85}.  Given
bivariate Gaussian initial conditions, the solution to the
Fokker-Planck equation at any time is a bivariate Gaussian and
can be computed in closed form. This Gaussian approximation is
accurate when $\thrX-\mu_\X\gg \sqrt{2D}$ for $\X=\subV,\subW$.

For simplicity, in this section we assume that the
two neurons receive statistically identical inputs so that $\mu_\subV=\mu_\subW=\mu$.
We further assume that the neurons are dynamically identical so that $g_\subV=g_\subW$, $V_{\text{rest}}=W_{\text{rest}}$,
and $\thrV=\thrW$.
The analysis is similar in the asymmetric case.
Without loss of generality, we rescale space so that $V_{\text{rest}}=W_{\text{rest}}=0$ and $\thrV=\thrW=1$.
To simplify calculations, we also time in units of the membrane time constants so that $g_\subV=g_\subW=1$.

For instance, the marginal or conditional firing rates can be approximated by the flux of the
time dependent  Gaussian distribution over threshold.  As shown in  Appendix
\ref{A:Gaussian}, this flux can be written in terms of the mean ($m$) and the variance ($\sigma^2$) of the Gaussian, and the input diffusion coefficient ($D$) as
\begin{equation}\label{E:nu}
J(m,\sigma^2,D):=\frac{(1-m)D}{\sqrt{2\pi}\sigma^3}e^{\frac{-(\mu-1)^2}{2\sigma^2}}.
\end{equation}                     
The steady state firing rate, $\nu_\infty$ is obtained by taking
$m=\mu$ and $\sigma^2=D$ to get $
\nu_\infty=J(\mu,D,D)=\frac{\alpha}{\sqrt{\pi}}e^{-\alpha^2}$ where $\alpha:=(1-\mu)\big / \sqrt{2 D}$. 
This expression can also be obtained using large deviation methods~\citep{vankampen,lindner01,SheaBrown:2008p1754}.

\begin{figure}
\begin{center}
\includegraphics[scale=0.5]{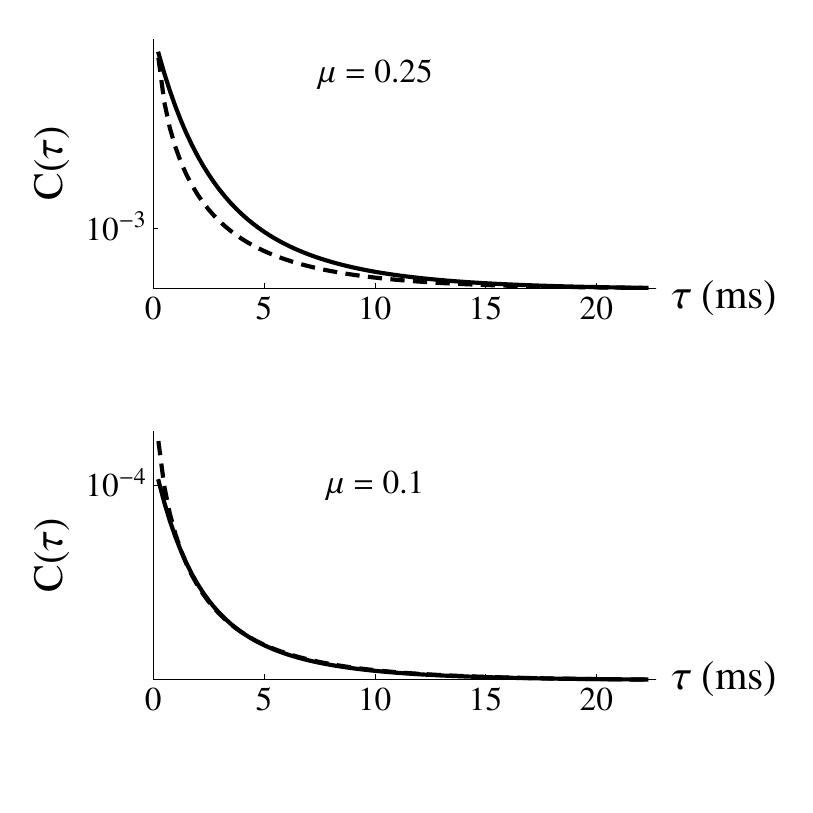}\hspace{.2in}
\includegraphics[scale=.75]{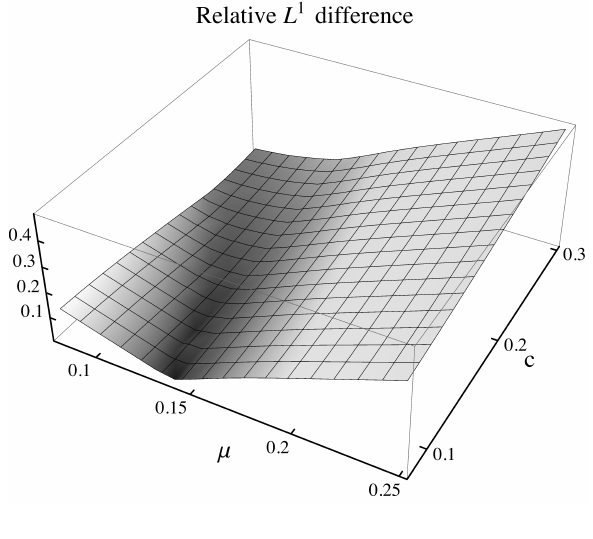}
\end{center}
\caption{Left: Cross-covariance functions for $\mu=0.1$ and
$\mu=0.25$ when $D=0.05$ and $c=0.2$.  The solid lines were obtained
from the Gaussian approximation and the dashed lines from finite
volume simulations. Right: The relative $L^1$ difference between the
Gaussian and finite volume cross-covariance functions ($L^1$
difference divided by the $L^1$ norm of the finite volume result)
for $\mu\in[0.075, 0.25]$ and $c\in[0.075, 0.3]$. The $L^1$ norm was
computed for $\tau>0.15$, due to the convergence issues of
$\nu_{V|W}$ for small $\tau$  discussed in Appendix
\ref{A:FVspikestats}.
The cross-covariance function has units Hz$^2$
In this figure and in Figs. \ref{F:Gnu} and \ref{F:GH}, the axes are labeled assuming a membrane time constant of $1/g_V=1/g_W=5$ms.
} \label{FigCorr}
\end{figure}

From an approximation of the conditional firing rate, the cross-covariance function can be obtained as (see Appendix \ref{A:Gaussian}),
\begin{align*}
C_{V|W}(\tau)=\nu_\infty(H(\tau)-\nu_\infty)
=\frac{1}{\pi }\alpha^2 e^{-\alpha ^2} \left(\frac{e^{t-\frac{\alpha ^2
   \left(e^\tau-c\right)}{c+e^\tau}}}{\sqrt{1-c^2 e^{-2 \tau}}
   \left(c+e^\tau\right)}-e^{-\alpha ^2}\right).
\end{align*}
In Fig.~\ref{FigCorr}, we compare this approximation to the cross-covariance function to the cross-covariance function obtained from finite volume simulations.  As expected, we find that the two agree well when firing rates and correlations are small, but disagree when $\mu$, $D$ or $c$ are larger.

Further expressions for other stationary and non-stationary spiking statistics
under the Gaussian approximation are derived in Appendix
\ref{A:Gaussian}.  We use these approximations to examine the response of
a pair of cells to time-varying inputs next.

\begin{figure}
\begin{center}
\includegraphics[scale=0.6]{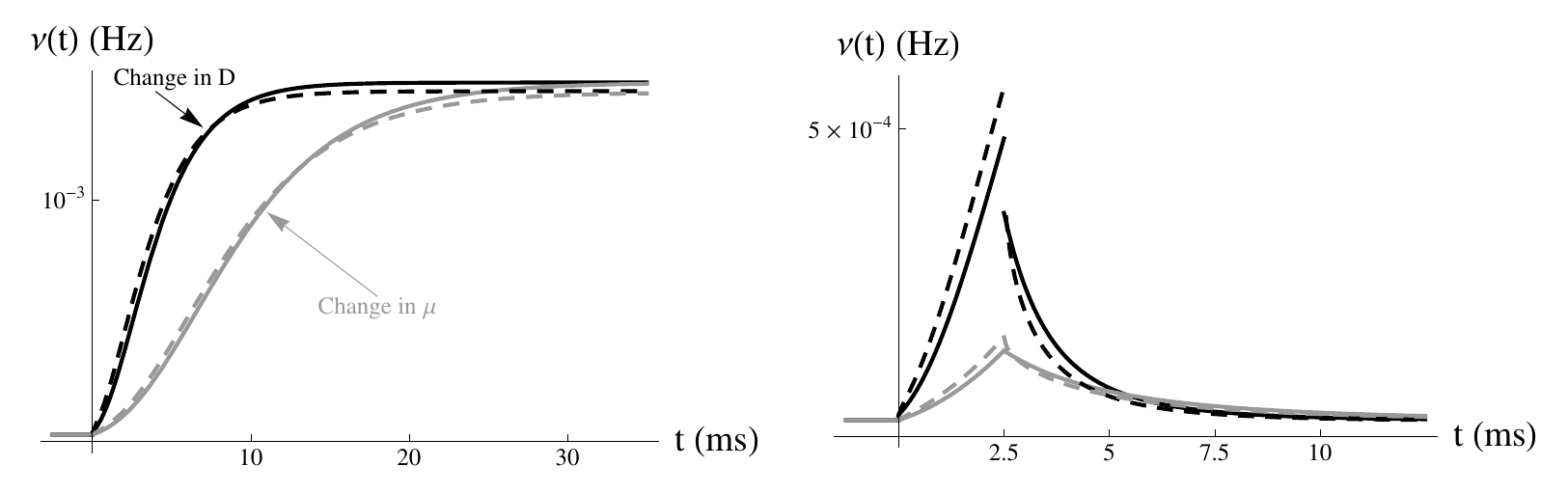}
\end{center}
\caption{\emph{Left}: Instantaneous firing rate after a step change in input statistics.  Result from the Gaussian approximation are plotted as solid lines and finite volume simulations as dashed lines.  For time $t<0$, the parameters were set to $\mu=0$ and $D=0.03$.  At time $t=0$ the parameters were changed to $\mu=0.1334$ (grey) or $D=0.04$ (black).  The values of $\mu$ and $D$ were chosen so that the steady state firing rate after the change in parameters was the same whether $\mu$ or $D$ was changed.
\emph{Right}: Instantaneous firing rate after a pulse change in input statistics. Same as Left, but the parameters were changed back to $\mu=0$ and $D=0.03$ at time $t=2.5$ms.
Note that a step change in the input variance, D, results in an instantaneous jump in the firing rate, followed by a continuous relaxation to the steady state.
To illustrate the quantitative accuracy of the Gaussian approximation, we used parameters that resulted in low firing rates.  Fig.~\ref{F:GH} illustrates that, while the Gaussian approximation is less accurate when firing rates are moderate, the approximation can still capture the qualitative behavior of the spiking statistics.
}
\label{F:Gnu}
\end{figure}

\subsection{Response to step changes in the input -- single cell response}

If a pair of cells responds rapidly to a change in an input
parameter, then the output of the cell pair can accurately capture
the information present in a time-varying input signal~\citep{Silberberg:2004p1474,Masuda:2006p129}.  It is therefore
useful to understand how the spiking statistics of a neuron or pair
of neurons respond to changes in input parameters, $\mu, D,$
and $c$ in Eq.~\eqref{E:model1}.
 The response of the cell pair is measured by their joint firing rate $(\nu_V, \nu_W)$, and we first examine how rapidly this response can track changes in the inputs to the model.

 We start by examining the response of a single cell.  Fig.~\ref{F:Gnu} \emph{Left} shows the time dependent firing rate after a step change in the mean, ($\mu$, light line), and variance, ($D$, heavy line).  We compare the response for the Gaussian approximation, derived in closed form in Appendix \ref{A:Gaussian}, to the result from finite volume simulations.
After a change in parameters, the distribution of $(V,W)$, and
therefore the spiking statistics, relax exponentially to a new
steady state.  However, the speed of this relaxation depends on the
parameters that are changed. After a step change in the mean input,
$\mu$, the firing rate relaxes to a new steady state with a time
constant of 1 (i.e., one membrane time constant).  After a change in
the variance, $D$, the firing rate jumps discontinuously, then
approaches the new steady state value with a faster time constant of
$1/2$ (see Appendix \ref{A:Gaussian}).  The fact that changes in the
variance of the input are tracked faster than changes in input
intensity is a fundamental property of the Ornstein-Uhlenbeck
process~\citep{gardiner85}.  Therefore, changes in variance can be
tracked more faithfully than changes in the DC component of the input.  Related observations are made in~\citep{Khorsand:2008p1757,Hasegawa,Silberberg:2004p1474,Masuda:2006p129}, where mainly the discontinuous change in output firing rate
in response to a step change in input variance was examined.

It also follows that a transient pulse change in $D$ results in a larger transient in the firing rate than a comparable pulse change in $\mu$. This prediction is verified for the Gaussian approximation and for finite volume simulations in Fig.~\ref{F:Gnu} \emph{Right}.

\subsection{Response to step changes  in the input -- joint response}

We next examine how joint response of the cell pair in response to a step
change in the input. The cross-covariance function, defined in Eq.~\eqref{E:CCGdef}, is commonly used to measure
dependencies between two spike trains over time.
Fig. \ref{FigCorr} compares the Gaussian approximation of the
cross-covariance function to that obtained using finite volume simulations. As expected,
the two results agree when firing rates are low. However, as $\mu$
or $D$ increase, the firing rate increases and the Gaussian approximation breaks
down.

\begin{figure}
\begin{center}
\includegraphics[scale=.7]{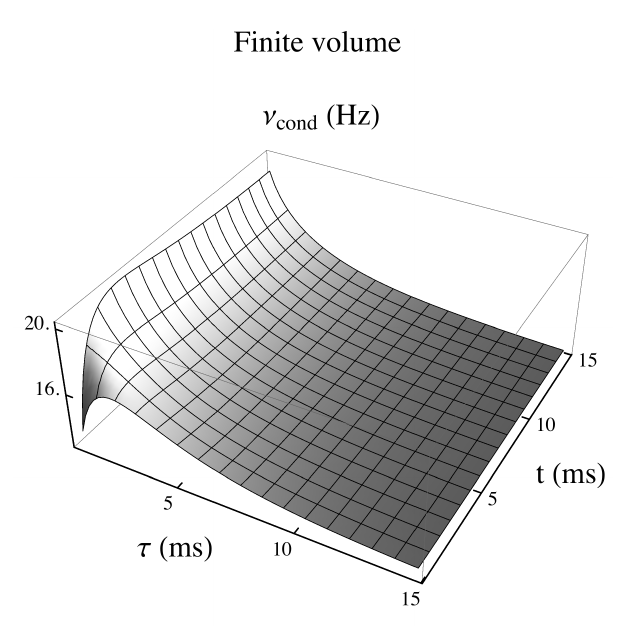}\hspace{.45in}\includegraphics[scale=.7]{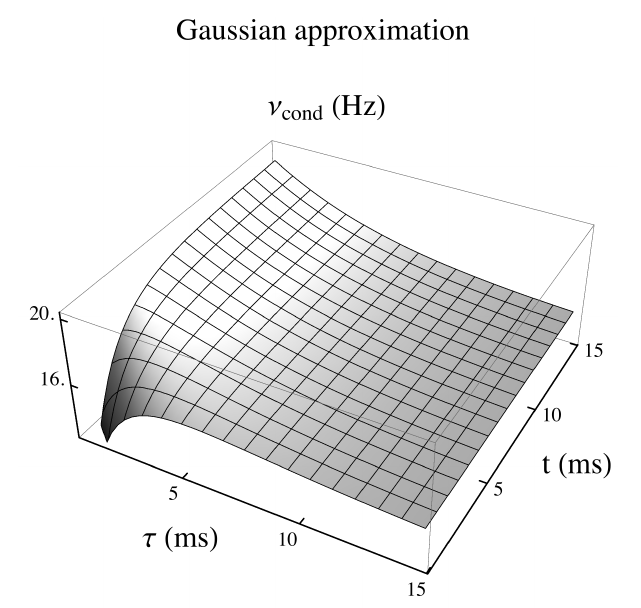}\\
\vspace{-.05in}
\includegraphics[scale=0.6]{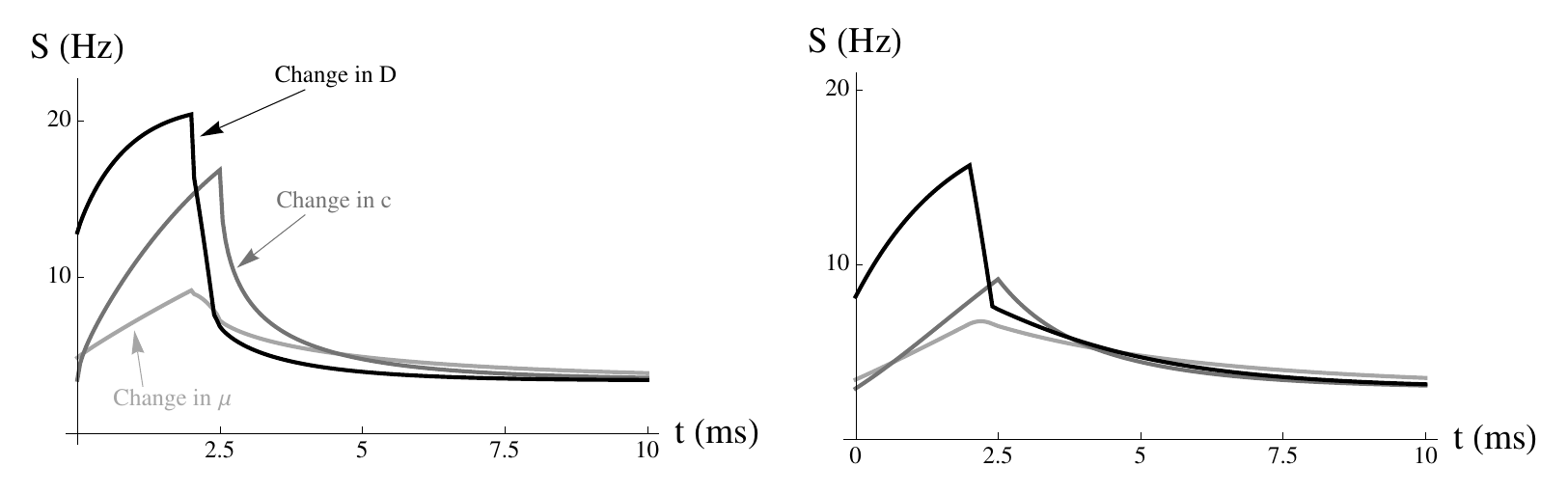}
\end{center}
\caption{\emph{Top row}: The two-point time dependent firing rate, $\nu_{\text{cond}}(\tau,t)$ after a step change in $D$ with $D=0.1$ for $t<0$ and $D=0.2$ for $t>0$. Parameters $\mu=0$ and $c=0.1$ were held constant.
Values for $\tau\le 0.5$ms were omitted due to the convergence issues discussed in Appendix \ref{A:FVspikestats}.
 \emph{Bottom row}: The conditional mean rate, $S$, after a pulse change in parameters.  The parameters changed from $D,c,\mu=0.1,0.1,0$ for $t<0$ to $D=0.2$ (black), $c=0.435$ (dark grey), or $\mu=0.293$ (light grey) for $t\in [0,2.5]$ms, then back to $D,c,\mu=0.1,0.1,0$ for $t>2.5$ms.
 The conditional mean rate, calculated according to Eq.~\eqref{E:En} with $a=0.15$ms and $b=0.5$ms, measures the propensity for cell $V$ to spike within the first $0.5$ms after cell $W$ spikes.  We chose $a=0.15$ms to circumvent the convergence issues discussed in Appendix \ref{A:FVspikestats}.
\emph{Left column}: finite volume simulations.  \emph{Right column}: Gaussian approximation.
 }
\label{F:GH}
\end{figure}

Fig.~\ref{F:GH} \emph{Top} shows the two-point conditional firing
rate after a step change in the parameter $D$. This function
completely characterizes the second order correlations of the two
cells over time. Such a plot would be computationally prohibitively
expensive to obtain using direct Monte-Carlo simulations, especially
when firing rates are low. The parameters for Fig.~\ref{F:GH} were
chosen so that the Gaussian approximation does not agree
quantitatively with the finite volume simulations. However, as shown
in Fig.~\ref{F:GH} \emph{Bottom}, the Gaussian approximation
successfully predicts the qualitative behavior of the bivariate
spiking statistics with changing input parameters.

A pulse change
in $D$ has a larger impact on the propensity of the cells to fire
together than a comparable pulse change in $\mu$ (compare to
Fig.~\ref{F:Gnu} \emph{Right}).  A pulse change in $c$ has an
intermediate impact.  If a downstream cell that receives inputs from
cells $V$ and $W$ is sensitive to synchrony in its
inputs~\citep{salinas00}, then the cell would response more quickly
and strongly to changes in $D$ or $c$ than to changes in $\mu$.

This suggests that downstream cells that act as coincidence detectors
are most sensitive to upstream changes in input variance, and respond more weakly
to changes in input intensity.

\section{Discussion}

Population density methods have a long history in neuroscience.  They have been
used to study both the statistics of responses of single neuron~\citep{tuckwell},  and
neural populations~\citep{harrison05}.
Studying the evolution of the ensembles,
rather than tracking individual neurons has several advantages:  While the dynamics
of each individual cell is stochastic, their probability density evolves deterministically.
The probability density is therefore easier to study analytically and numerically.
For instance, both linear response methods~\citep{lindner01,rocha07,ostojic}, and
the Gaussian approximation discussed here (see also~\citep{burak09,tchumatchenko10,badel10}), are
obtained by considering the evolution of large populations in the diffusive limit.
Simulating the evolution of population densities is typically orders of magnitudes
faster than simulating the evolution of each individual cell in a population~\citep{Nykamp00,Omurtag00}.  For instance, obtaining the two--point time dependent
firing rate, $\nu_{\text{cond}}(\tau,t)$ shown in Fig.~\ref{F:GH}, would not be feasible
using Monte Carlo methods on an average machine today.

We have concentrated on a simple version of the model to keep the presentation relatively concise.  For instance, it is easy to consider cells receiving different, potentially
time dependent drives, $\mu_V(t), \mu_W(t), D_V(t)$ and $D_W(t)$.
We have mainly considered drift terms of the form, $f(V,W) = f(V)$, and $g(W,V) = g(W)$,  in Eq.~\eqref{E:model1}, so that the two populations were uncoupled.
Coupled populations have been considered earlier~\citep{Nykamp00}.  We hence concentrated on examining the effects of anisotropic diffusion.  However, the numerical methods  we described can easily handle coupling between cells in the sub--threshold regime.
This could be used to examine the interplay of correlated inputs and cell coupling~\citep{Schneider:2006p254}.  However, we do not know whether there is a direct way to include super-threshold coupling in the present diffusive approximation.   We also note
that high firing rates can lead to steep gradients of the probability density close to the boundary and convergence problems in the numerical methods we used.  This suggests that numerical techniques
will have to be developed further to accurately capture the response of IF neurons
in this regime.

We have used our numerical and analytical approach to examine the best
way to transmit information in a pair of cells.   We found that both the single
cell response and the joint response tracks changes in noise intensity more accurately
than changes in the mean drive or correlations in the cell inputs.   Thus 
input variance appears to provide the best channel to code information at the single cell
and population level. 

 The numerical 
methods we have developed can be used to further examine how the output of a
cell pair reflects their interactions and dependencies in their inputs.   It is of
particular interest how cells respond to signals that vary in time.  The finite
volume method we described is well suited to this task, as it is designed to capture the time--dependent response
of a cell pair. 

\appendix

\section{Numerical schemes}
\label{numerical}

In the following we provide a description of the finite volume method used in the numerical
simulations.   Some of the details of the implementation are not standard.  As we are not
aware of a similar treatment of this type of equation, we give a detailed discussion of the novel
aspects of the algorithm.

\begin{figure*}
\begin{center}
\includegraphics[width=0.7\textwidth]{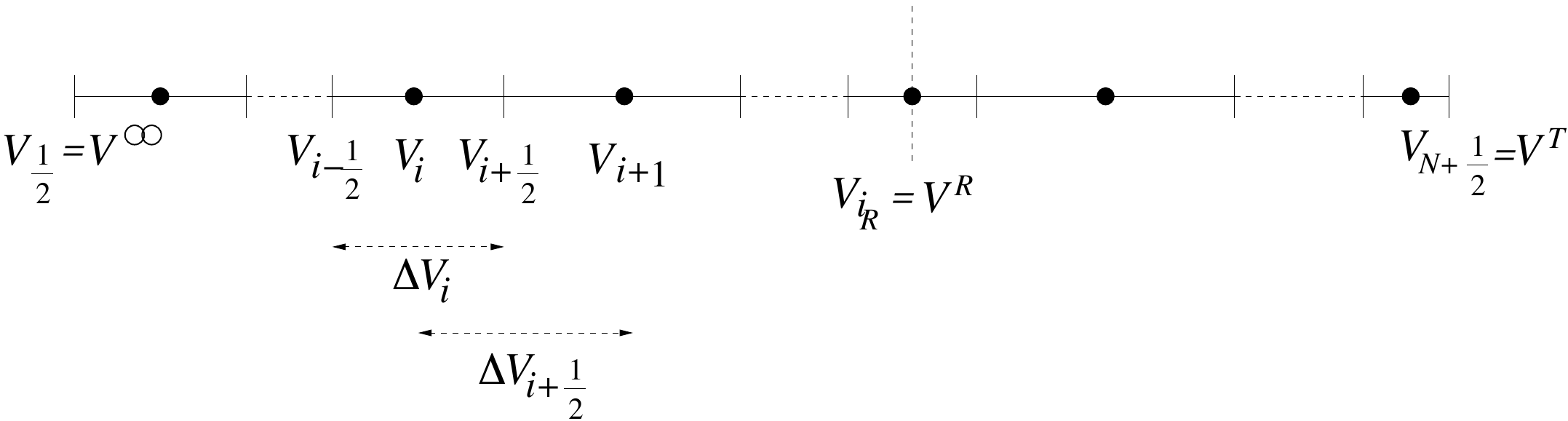}
\end{center}
\caption{A schematic depiction of the subdivision of the domain $(\vinf ,\vtt )$ into subintervals $(\vv _{i-\ff },\vv _{i+\ff })$. The reset voltage $\vrr $ is at $\vv _{\ir }$. Similar notations are used for the second neuron voltage $W$.
}
\label{fig:mesh1d}       
\end{figure*}

The time steps are defined by $\Delta t_n=t_{n+1}-t_{n}$. When there is no ambiguity, the time step is denoted by $\Delta t$.
The intervals $(\vinf ,\vtt )$ and $(\winf ,\wtt )$ are partitioned into $N_{\vv }$ and $N_{\ww }$ sub-intervals respectively. We denote $\Delta \vv _{i}$ as the $i^{th}$ step in the $\vv $-direction and $\Delta \ww _{j}$ as the $j^{th}$ step in the $\ww $-direction,
for $i=1,\ldots, N_{\vv }$, $j=1,\ldots, N_{\ww }$. Our quadrilateral mesh elements, $Q_{i,j}$, are then defined by
$$Q_{i,j}=(\vv _{i-\frac 1 2},\vv _{i+\frac 1 2})\times (\ww _{j-\frac 1 2},\ww _{j+\frac 1 2}),$$
with $\vv _{i-\frac 1 2}=\sum_{l=1}^{i-1}\Delta \vv _{l}$, $\ww _{j-\frac 1 2}=\sum_{l=1}^{j-1}\Delta \ww _{l}$. Our mesh-points $(\vv _{i},\ww _{j})$ are the centroids of the cells, thus $\vv _{i}=\vv _{i-\ff }+\frac{1}{2}\Delta \vv _{i}$, $\ww _{j}=\ww _{j-\ff }+\frac{1}{2}\Delta \ww _{j}$. We make sure that there exist two indices $i_{R}$ and $j_{R}$ such that $\vv _{\ir }=\vrr $ and $\ww _{\ir }=\wrr $, which means that the two reset potentials fall exactly on some mesh-points, see Fig.~\ref{fig:mesh1d}.
For every function $\xi $ defined on $(0,T)\times \Omega $, the notation $\xi _{\alpha, \beta }^n$ stands for the approximation of $\xi \big( t_{n},(\vv _{\alpha }, \ww _{\beta })\big)$, for $\alpha =i, i\pm \ff $, $\beta =j, j\pm \ff $. We denote $\xi ^n$ as the sequence $\{ \xi _{i,j}^n\} _{i,j}$.

\subsection{Treatment of the drift operator: scheme $\mathcal{S}_{1}$}
\label{subsection:drift} The advection equation in
(\ref{eq:transport}) is discretized by using the one-dimensional
numerical fluxes in \citep{faj_neuro} in each direction. We set
\begin{equation}
P_{i,j}^{n+1}=P_{i,j}^{n}-\frac{\Delta t}{\Delta v_{i}}(\mathcal{A}_{i+\ff ,j}^{n}-\mathcal{A}_{i-\ff ,j}^{n})
-\frac{\Delta t}{\Delta w_{j}}(\mathcal{A}_{i,j+\ff }^{n}-\mathcal{A}_{i,j-\ff }^{n}),\label{eq:limited2d}
\end{equation}
where the numerical fluxes are defined by
\begin{multline}
\mathcal{A}_{i+\ff ,j}^{n}=f_{i+\ff ,j}^{+}P_{i,j}^{n}+f_{i+\ff ,j}^{-}P_{i+1,j}^{n}\\
+ \frac{1}{2}\frac{\Delta P_{i+\ff ,j}^{n}}{\Delta v_{i+\ff }}\Big( f_{i+\ff ,j}^{+}\Delta v_{i} \varphi \big( r_{i+\ff ,j}^{p},\frac{\Delta v_{i+\ff }}{\Delta v_{i}}\big) - f_{i+\ff ,j}^{-}\Delta v_{i+1}\varphi \big( r_{i+\ff ,j}^{m},\frac{\Delta v_{i+\ff }}{\Delta v_{i+1}}\big) \Big) ,\label{eq:vflux2d}
\end{multline}
for $1\le i\le N_{\vv }-1$, $1\le j\le N_{\ww }$ and
\begin{multline}
\mathcal{A}_{i,j+\ff }^{n}=g_{i,j+\ff }^{+}P_{i,j}^{n}+g_{i,j+\ff }^{-}P_{i,j+1}^{n}\\
+ \frac{1}{2}\frac{\Delta P_{i,j+\ff }^{n}}{\Delta w_{i,j+\ff }}\Big( g_{i,j+\ff }^{+}\Delta w_{j}\varphi \big( r_{i,j+\ff }^{p},\frac{\Delta w_{j+\ff }}{\Delta w_{j}}\big) - g_{i,j+\ff }^{-}\Delta w_{j+1}\varphi \big( r_{i,j+\ff }^{m},\frac{\Delta w_{j+\ff }}{\Delta w_{j+1}}\big)\Big) ,\label{eq:wflux2d}
\end{multline}
for $1\le i\le N_{\vv }$, $1\le j\le N_{\ww }-1$,
with notations $\Delta P_{i+\ff ,j}^{n}:=P_{i+1,j}^{n}-P_{i,j}^{n}$, $\Delta P_{i,j+\ff }^{n}:=P_{i,j+1}^{n}-P_{i,j}^{n}$,
$$\ds{r_{i+\ff ,j}^{p}=\frac{f_{i-\ff ,j}^{+}\Delta P_{i-\ff ,j}}{f_{i+\ff ,j}^{+} \Delta P_{i+\ff ,j}},\ r_{i-\ff ,j}^{m}=\frac{f_{i+\ff ,j}^{-}\Delta P_{i+\ff ,j}}{f_{i-\ff ,j}^{-} \Delta P_{i-\ff ,j}}\ ,} \ds{r_{i,j+\ff }^{p}=\frac{g_{i,j-\ff }^{+}\Delta P_{i,j-\ff }}{g_{i,j+\ff }^{+} \Delta P_{i,j+\ff }},\ r_{i,j-\ff }^{m}=\frac{g_{i,j+\ff }^{-}\Delta P_{i,j+\ff }}{g_{i,j-\ff }^{-} \Delta P_{i,j-\ff }}\ }\ ,$$
where the limiter function $\varphi $ is defined by $\varphi (a,b)=2b\max \big( 0,\min(1,2a),\min(a,2)\big)$.

To comply with Dirichlet boundary conditions in (\ref{eq:transport}), we further impose $\mathcal{A}_{0,j}^{n}=\mathcal{A}_{N_{\vv}+\ff ,j}^{n}=\mathcal{A}_{i,0}^{n}=\mathcal{A}_{i,N_{\ww }+\ff }^{n}=0$, for $1\le i\le N_{\vv }$, $1\le j\le N_{\ww }$.
\begin{proposition}
The numerical scheme (\ref{eq:limited2d}) is non-negativity preserving under the {\sc{cfl}} condition
\begin{equation*}
\Delta t \Big( \frac{f _{i-\ff ,j}^{+}-f_{i+\ff ,j}^{-}}{\Delta v_{i}}+\frac{g_{i,j-\ff }^{+}-g_{i,j+\ff }^{-}}{\Delta w_{j}}
+ \Big( \frac{f_{i+\ff ,j}-f_{i-\ff ,j}}{\Delta v_{i}}\Big)^{+}+\Big(\frac{g_{i,j+\ff }-g_{i,j-\ff }}{\Delta w_{j}}\Big) \Big)^{+} \le 1\ .
\end{equation*}
\end{proposition}

The proof of this proposition is similar to the proof of the
corresponding one-dimensional result in \citep{faj_neuro}, and we
therefore omit it here.

\subsection{Treatment of the diffusion operator: scheme $\mathcal{S}_{2}$}
\label{subsection:diffusion}
Our approximation of the solutions to (\ref{eq:diffreac}) in two space dimensions is given by the implicit scheme
\begin{equation}
\frac{\Delta v_{i}\Delta w_{j}}{\Delta t}P_{i,j}^{n+1} - (\mathcal{B}_{i+\ff ,j}^{n+1}-\mathcal{B}_{i-\ff ,j}^{n+1}) \\
- (\mathcal{B}_{i,j+\ff }^{n+1}-\mathcal{B}_{i,j-\ff }^{n+1})=\frac{\Delta v_{i}\Delta w_{j}}{\Delta t}P_{i,j}^{n} + \delta _{i,i_{R}}S^{\ww ,n+1}_{j}+\delta _{j,j_{R}}S^{\vv ,n+1}_{i},\ \label{eq:schemediff2d}
\end{equation}
where $\delta _{k_{1},k_{2}}$ is the symbol of Kronecker and $S_{i}^{\vv ,n+1}$ and $S_{j}^{\ww ,n+1}$ account for the re-injection condition (\ref{eq:bc2d6}), see subsection \ref{subsection:bc}. The numerical diffusive fluxes are defined by
\begin{equation}
\mathcal{B}_{i+\ff ,j}^{n+1}=D\frac{P_{i+1,j}^{n+1}-P_{i,j}^{n+1}}{\Delta v_{i+\ff }}\\
+\frac{CD}{2}\Big( \frac{P_{i+1,j+1}^{n+1}-P_{i+1,j}^{n+1}}{\Delta w_{j+\ff }}+\frac{P_{i,j}^{n+1}-P_{i,j-1}^{n+1}}{\Delta w_{j-\ff }}\Big)\ , \label{eq:vfluxdiff2d}
\end{equation}
for $1\le i\le N_{\vv }-1$, $1\le j\le N_{\ww }$ and
\begin{equation}
\mathcal{B}_{i,j+\ff }^{n+1}=D\frac{P_{i,j+1}^{n+1}-P_{i,j}^{n+1}}{\Delta w_{j+\ff }}\\
+\frac{CD}{2}\Big( \frac{P_{i+1,j+1}^{n+1}-P_{i,j+1}^{n+1}}{\Delta v_{i+\ff }}+\frac{P_{i,j}^{n+1}-P_{i-1,j}^{n+1}}{\Delta v_{i-\ff }}\Big)\ ,
\label{eq:wfluxdiff2d}
\end{equation}
for $1\le i\le N_{\vv }$, $1\le j\le N_{\ww }-1$.
\begin{remark}
Notice that the second term of the right-hand side in
(\ref{eq:vfluxdiff2d}) stand for a centered finite difference
discretization of the cross derivative $CD\partial ^{2}_{vw}P$ on
the right vertical interface of $Q_{i,j}$. Other numerical schemes
have been implemented in \citep{fmchbms,patrick,bkst}, but yield
unconditionally positive off-diagonal coefficients in the diffusion
matrix, therefore producing negative undershoot near sharp solution
gradients. When the neurons are strongly correlated (i.e.: when
$C\thickapprox 1$), the gradients of the solution can be very sharp.
The advantage of our method is that all the off-diagonal
coefficients are nonnegative where the mesh is uniform, which means,
the region where the solution is not $0$ in practice. Using a
similar remark in (\ref{eq:wfluxdiff2d}), the resulting numerical
scheme (\ref{eq:schemediff2d}) is nonnegative in realistic
applications.
\end{remark}

The boundary conditions in (\ref{eq:diffreac}) are implemented as $\mathcal{B}_{\ff ,j}^{n+1}=\mathcal{B}_{i,\ff }^{n+1}=0$ and
$$\mathcal{B}_{N_{\vv }+\ff,j}^{n+1}=D\frac{0-P_{N_{\vv },j}^{n+1}}{\Delta \vv _{N_{\vv }/2}}\ ,\mathcal{B}_{i,N_{\ww }+\ff }^{n+1}=D\frac{0-P_{i,N_{\ww }}^{n+1}}{\Delta \ww _{N_{\ww }+\ff }}\ ,$$
\begin{equation*}
\mathcal{B}_{i+\ff ,N_{\ww }}^{n+1}=D\frac{P_{i+1,N_{\ww }}^{n+1}-P_{i,N_{\ww }}^{n+1}}{\Delta \vv _{i+\ff }}
+\frac{CD}{2}\Big( \frac{0-P_{i+1,N_{\ww }}^{n+1}}{\Delta \ww _{N_{\ww }}/2}+\frac{P_{i,N_{\ww }}^{n+1}-P_{i,N_{\ww }-1}^{n+1}}{\Delta \ww _{N_{\ww }-\ff }}\Big)\ ,
\end{equation*}
\begin{equation*}
\mathcal{B}_{N_{\vv },j+\ff }^{n+1}=D\frac{P_{N_{\vv },j+1}^{n+1}-P_{N_{\vv },j}^{n+1}}{\Delta \ww _{j+\ff }}
+\frac{CD}{2}\Big( \frac{0-P_{N_{\vv },j+1}^{n+1}}{\Delta \vv _{N_{\vv }}/2}+\frac{P_{N_{\vv },j}^{n+1}-P_{N_{\vv }-1,j}^{n+1}}{\Delta \vv _{N_{\vv }-\ff }}\Big)\ .
\end{equation*}

\subsection{Treatment of the re-injection condition (\ref{eq:bc2d6})}
\label{subsection:bc}
\begin{figure*}
\begin{center}
\includegraphics[width=0.7\textwidth]{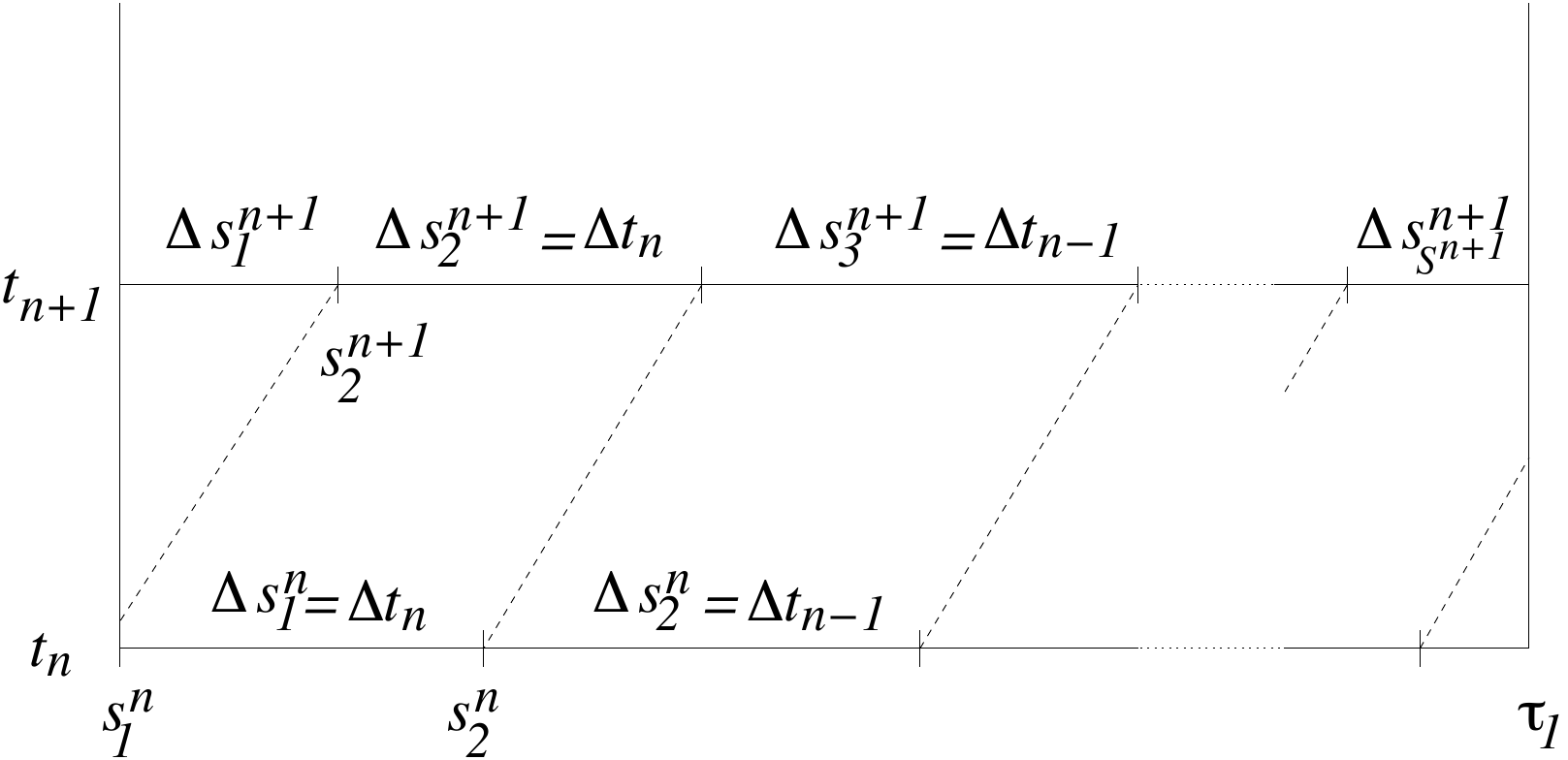}
\end{center}
\caption{A schematic depiction of the time dependent subdivision of the domain $(0,\tau _{\vv })$ into subintervals $(s^{n}_{k},s^{n}_{k+1})$.
}
\label{figure:age}       
\end{figure*}

Since the time variable $t$ and the age variable $s\in (0,\tau _{1})$ evolve together, the domain $(0,\tau _{1})$ is dynamically partitioned into sub-intervals $(s^{n}_{k},s^{n}_{k+1})$ such that $s^{n}_{k}=0$ and $s^{n}_{k+1}=\min( s^{n}_{k}+\Delta s_{k}^{n},\tau _{1})$, where the age steps $\Delta s_{k}^{n}$ match the time steps as follows: for all $n$, $\Delta s_{1}^{n}=\Delta t_{n}$ and $\Delta s_{k}^{n}=\Delta t_{n-k}$, $k=1,\ldots ,n-1$. We set $K^{s}_{n}=\max \{k,\ s^{n}_{k}<\tau _{1}\}$, see figure \ref{figure:age}. In the same way, we discretize $(0,\tau _{2})$ into sub-intervals $(r^{n}_{k},r^{n}_{k+1})$ such that $r^{n}_{k}=0$ and $r^{n}_{k+1}=\min( r^{n}_{k}+\Delta r_{k}^{n},\tau _{1})$, with $\Delta r_{1}^{n}=\Delta t_{n}$, $\Delta r_{k}^{n}=\Delta t_{n-k}$, $k=1,\ldots ,n-1$. We set $K^{r}_{n}=\max \{k,\ r^{n}_{k}<\tau _{2}\}$.

Then, defining the piecewise constant functions
\begin{equation}
\tr _{1,i}(t)= \sum _{n=0}^{+\infty } R_{1,i}^{n+1,K^{r}_{n}}\chi _{(t_{n},t_{n+1}]}(t)\  ,\ \tr _{2,j}(t)= \sum _{n=0}^{+\infty } R_{2,j}^{n+1,K^{s}_{n}}\chi _{(t_{n},t_{n+1}]}(t)\ ,
\nonumber 
\end{equation}
the quantities
$$S^{\vv }_{i}:=\int _{\tau _{2}-\Delta t_{n}}^{\tau _{2}}\tr _{1,i}(t)\, dt \textrm{ and } S^{\ww }_{j}:=\int _{\tau _{1}-\Delta t_{n}}^{\tau _{1}}\tr _{2,j}(t)\, dt$$
are injected into (\ref{eq:schemediff2d}).

\subsection{Treatment of the one neuron boundary condition (\ref{eq:bc2d2})--(\ref{eq:bc2d5})}
\label{subsection:1dfp} Equation (\ref{eq:bc2d2}) is discretized
with the one--dimensional numerical scheme in \citep{faj_neuro}. We
use the operator splitting technique
\begin{equation}
R_{1,i}^{n+\ff ,k+\ff }=R_{1,i}^{n,k}-\frac{\Delta t}{\Delta V_{i}}(\mathcal{A}_{i+\ff }^{n}-\mathcal{A}_{i-\ff }^{n})\ ,\textrm{ and } R_{1,i}^{n+1 ,k+1 }-\frac{\Delta t}{\Delta V_{i}}(\mathcal{B}_{i+\ff }^{n+1}-\mathcal{B}_{i-\ff }^{n+1}) = R_{1,i}^{n+\ff ,k+\ff }+\delta _{i,i_{R}}S^{\ww ,n+1},
\label{eq:splitting1d}
\end{equation}
where the advective and diffusive numerical fluxes are
\begin{equation}
\mathcal{A}_{i+\ff }^{n}=f_{i+\ff ,N_{\ww }}^{+}R_{1,i}^{n,k}+f_{i+\ff ,N_{\ww }}^{-}R_{1,i+1}^{n,k} + \frac{1}{2}\frac{\Delta R_{1,i+\ff}^{n,k}}{\Delta \vv _{i+\ff }}\Big( f_{i+\ff ,N_{\ww }}^{+}\Delta \vv _{i} \varphi \big( r_{i+\ff }^{p},\frac{\Delta v_{i+\ff }}{\Delta v_{i}}\big) - f_{i+\ff ,N_{\ww }}^{-}\Delta \vv _{i+1}\varphi \big( r_{i+\ff }^{m},\frac{\Delta v_{i+\ff }}{\Delta v_{i+1}}\big)\Big) ,\nonumber 
\end{equation}
\begin{equation}
\mathcal{B}_{i+\ff }^{n+1}=D\frac{R_{1,i+1}^{n+1,k+1}-R_{1,i}^{n+1,k+1}}{\Delta v_{i+\ff }}
\end{equation}
for $i=1,N_{\vv }-1$, and $\mathcal{A}_{\ff }^{n}=\mathcal{A}_{N_{\vv }+\ff }^{n}=\mathcal{B}_{\ff }^{n+1}=0$, $\mathcal{B}_{N_{\vv }+\ff }^{n+1}=-\frac{D}{2\Delta \vv _{N_{\vv }}}R_{1,N_{\vv }}^{n+1,k+1}$.

We use similar arguments to discretize Eq.~(\ref{eq:bc2d2}).
Then, since the age steps $\Delta s^{n}_{k}$ match the time steps $\Delta t_{n-k}$, equation (\ref{eq:bc2d4}) is solved exactly as $R^{n+1,k+1,l+1}=R^{n,k,l}$, and its boundary conditions (\ref{eq:bc2d1prime}) translate into $R^{n,0,l}=-\frac{D}{2\Delta \vv _{N_{\vv }}}R_{1,N_{\vv }}^{n,l}$, $R^{n,k,0}=-\frac{D}{2\Delta \ww _{N_{\ww }}}R_{2,N_{\ww }}^{n,k}$.

Finally, the first re--injection condition in (\ref{eq:bc2d5}) is taken into account by setting $S^{\ww ,n+1,l}:=R^{n+1,K^{s},l}$ in (\ref{eq:splitting1d}). The second re--injection condition in (\ref{eq:bc2d5}) is discretized in a similar fashion.

\subsection{Comparison with solutions obtained using Monte Carlo methods}
\label{MC}

Monte Carlo simulations were performed by integrating the Langevin
equations ($\ref{E:model1}$) using the Euler--Maruyama method with
time step $10^{-3}$~\citep{kloeden}.   Histograms in the $V-W$ plane
were typically created using $2\times 10^{7}$ points sampled from a
long, simulated trajectory and using the same grid on the domain
$\Omega = (\vinf,\vtt )\times (\winf ,\wtt )$  as in the finite
volume simulation.  To compare the results of the finite volume and
Monte Carlo methods, we computed the $L^{1}$ norm of their
difference by taking the absolute value of the difference at each
grid point and summing over the domain $\Omega$.

\subsection{Calculating spike train statistics from finite volume simulations}\label{A:FVspikestats}

To obtain spike train statistics from finite volume simulations, we
used the definitions from Sec.~\ref{S:spikestats}.  We solved the
two-dimensional Fokker-Planck equation using the numerical scheme
described above.  To solve the one-dimensional Fokker-Planck equation
(for example, to obtain instantaneous or conditional firing rates
\emph{cf.} Eq.~\eqref{E:F1d}), we used the one-dimensional scheme
described in \citep{faj_neuro} or we used the two-dimensional scheme
and calculated the marginal density from the two-dimensional density
as $P_1(t,V)=\int_{W^{-\infty}}^{W^T}P(t,V,w)dw$.

However, we met with convergence problems in calculating the conditional
flux $J_{\text{cond}}(t,V)$ using Eq.~\eqref{CdlFlux}.  These are due 
to the fact that when $W$ is
close to the threshold value $W^{T}$, and the two neurons fire nearly synchronously,
the re-injection process  in the $V$ and $W$ directions is relatively complicated.
Given these
convergence issues, we ignored the first $0.5$ms in the left panel of
Fig.~\ref{FigCorr} and $0.75$ms in right panel of Fig.~\ref{FigCorr} and Fig.~\ref{F:GH} when we computed the
conditional firing rate $\nu_{V|W}(\tau,t)$.


\section{Gaussian approximation of the LIF}\label{A:Gaussian}

In this section we derive an approximation of the spiking statistics for the LIF in low firing rate regimes.
Recall that the LIF is defined by taking $f(V,W)=-g_V (V - V_{\text{rest}})$ and $g(V,W) = -g_W (W - W_{\text{rest}})$ in Eq.~\eqref{E:model1}.
For simplicity, in this section we assume that the
two neurons receive statistically identical inputs so that $\mu_\subV=\mu_\subW=\mu$.
We further assume that the neurons are dynamically identical so that $g_\subV=g_\subW$, $V_{\text{rest}}=W_{\text{rest}}$,
and $\thrV=\thrW$.
The analysis is similar in the asymmetric case.
Without loss of generality, we rescale space so that $V_{\text{rest}}=W_{\text{rest}}=0$ and $\thrV=\thrW=1$.
To simplify calculations, we also time in units of the membrane time constants so that $g_\subV=g_\subW=1$.

When $\alpha:=(1-\mu)\big / \sqrt{2 D}$ is large, firing rates are low and the boundary conditions at threshold have a small impact on the distribution.  Also, since the cells only spike rarely, the refractory period can be ignored.
In such regimes, the solution of the full problem is approximated by the solution of the free boundary problem.
This approximation, which we call the Gaussian approximation, is accurate in the limit  $\alpha\to \infty$.
In this case the membrane potentials $(V(t),W(t))$ are described by an Ornstein-Uhlenbeck process on $\Rr^2$.  Such processes are well-understood and the spiking statistics can be computed exactly, as we show below.

Assume that the initial distribution $P(0,U)$ is a bivariate Gaussian with marginal means $m(0)=E[V(0)]=E[W(0)]$, variance $\sigma^2(0)=\var(V(0))=\var(W(0))$, and covariance $\gamma(0)=\cov(V(0),W(0))$. Then the
solution at any time $t\ge 0$ (in the absence of boundary conditions) is Gaussian with mean, variance and covariance given respectively by
$$
\begin{array}{lcl}
m(t)&=&e^{-t}m(0)+\left(1-e^{-t}\right)m(\infty),\medskip\\
\sigma^2(t)&=&e^{-2t}\sigma^2(0)+\left(1-e^{-2t}\right)\sigma^2(\infty),\text{ and}\medskip\\
\gamma(t)&=&e^{-2t}\gamma(0)+\left(1-e^{-2t}\right)\gamma(\infty)
\end{array}
$$
where
$$
\begin{array}{lcl}
m(\infty)=\mu,\quad
\sigma^2(\infty)=D,\quad\text{and}\quad
\gamma(\infty)=cD
\end{array}
$$
are the steady state mean, variance, and covariance.

\subsection{Conditional firing rate and spike count correlation in the steady state}

The results above can be used to derive an approximation of the steady state conditional firing rates.
Since the joint distribution of $(\V,\W)$ is a bivariate Gaussian, the distribution of $\V$ given that $\W=\thrW=1$ is a univariate Gaussian.  The conditional mean and variance are
$
m_c(0)=c(1-\mu)+\mu
$
and
$
\sigma_c^2(0)=D(1-c^2)
$
respectively.  As time evolves, the conditional density of $\V$ relaxes to its steady state.  The density during this relaxation is a univariate Gaussian with mean
$$
m_c(\tau)=e^{-\tau}m_c(0)+\left(1-e^{-\tau}\right)m_c(\infty)
$$
and variance
$$
\sigma_c^2(\tau)=e^{-2\tau}\sigma^2_c(0)+\left(1-e^{-2\tau}\right)\sigma^2_c(\infty)
$$
where $m_c(\infty)=\mu$ and $\sigma^2_c(\infty)=D$ are the stationary mean and variance.
Note that for $c$ near 1, $m_c(0)$ is near $\thrV=1$ which violates the assumptions of the Gaussian approximation, namely that the mass near threshold is small.  In this case, the \emph{conditional} flux across threshold is large even if the \emph{marginal} fluxes across threshold are small.  Thus, for the approximation of the conditional firing rate to be accurate, we must assume that $\alpha$ is large \emph{and} that $c$ is small.

The conditional firing rate is simply
$$
\nu_{V|W}(\tau)=J(\mu_c(\tau),\sigma^2_c(\tau),D)
$$
where $J(\mu,\sigma^2,D)$ is as defined in \eqref{E:nu}.
This expression does not depend on $t$ because we have assumed that the two-dimensional distribution is in its steady state.
Later, we look at the two point conditional firing rate outside of the steady state.

The steady state cross-covariance function is obtained from the conditional firing rate \emph{cf.} Eq.~\eqref{E:CCGdef} to obtain
\begin{align*}
C_{V|W}(\tau)=\nu_\infty(H(\tau)-\nu_\infty)
=\frac{1}{\pi }\alpha^2 e^{-\alpha ^2} \left(\frac{e^{t-\frac{\alpha ^2
   \left(e^\tau-c\right)}{c+e^\tau}}}{\sqrt{1-c^2 e^{-2 \tau}}
   \left(c+e^\tau\right)}-e^{-\alpha ^2}\right)
\end{align*}
To first order in $c$ this gives,
\begin{equation}\label{E:ccgc1}
\CCG(\tau)=\frac{c}{\pi }\alpha ^2 \left(2 \alpha ^2-1\right) e^{-2 \alpha
   ^2-\tau}+o(c^2).
\end{equation}
The asymptotic spike count correlation, defined by
$$
\rho:=\lim_{t\to\infty}\frac{\cov(N_V(t),N_W(t))}{\sqrt{\var(N_V(t))\var(N_W(t))}},
$$
can be written in terms of the conditional firing rate
as~\citep{SheaBrown:2008p1754}
$$
\rho=\frac{2\int_{0}^\infty \left(\nu_{V|W}(\tau)-\nu_\infty\right)d\tau}{CV^2}
$$
where $CV$ is the coefficient of variation of the spike train inter-spike intervals.  In the low firing rate, $\alpha\to \infty$ limit, $CV\to 1$ and therefore, to first order in $c$ and $\nu_\infty$,
\begin{align}
\rho&\approx 2\int_{0}^\infty \left(\nu_{V|W}(\tau)-\nu_\infty\right)d\tau\notag\\
&=\frac{c}{\sqrt{\pi }}2 \alpha  \left(2 \alpha ^2-1\right)  e^{-\alpha^2}.\label{E:rhoc1}
\end{align}
Since both $\nu_\infty$ and the correlation susceptibility,
$T:=\frac{\rho}{c}$, are functions of the single parameter $\alpha$
and since $\nu_\infty$ is monotonic with $\alpha$, we may conclude
that $T$ is a function of $\nu_\infty$ to first order in $c$ and
$\nu_\infty$.  This same conclusion was reached
in~\cite{SheaBrown:2008p1754} using linear response theory, though
the expression derived for $\rho$,
\begin{equation}\label{E:rhoPRL} 
\rho\approx\frac{c}{\sqrt{\pi }}\alpha  \left(2 \alpha -\frac{1}{\alpha }\right)^2,
   e^{-\alpha ^2}
\end{equation}
differs from Eq.~\eqref{E:rhoc1}.  For both expressions,
$\frac{\partial \rho}{\partial \nu_\infty}\sim 4c\alpha^2$ as
$\alpha\to \infty$ (i.e., as $\nu_\infty\to 0$). Comparing these two
approximations to a more accurate linear response approximation
(also from~\citep{SheaBrown:2008p1754}), we found that
Eq.~\eqref{E:rhoPRL} is more accurate than Eq.~\eqref{E:rhoc1}.

\subsection{Time dependent input statistics}

We will now investigate how the spiking statistics track time
dependent changes in the inputs. When the input parameters to the
neurons are time dependent, the two dimensional density $p(t,V,W)$ at
any time $t$ is a bivariate Gaussian whenever the initial condition
is a bivariate Gaussian. Thus we can use the same methods as above
to derive the time dependent spiking statistics. To illustrate the
effects of time-dependent inputs, we concentrate on a simple
time-dependent input model. We assume that each cell receives input
with mean $\mu_0$, diffusion $D_0$, and correlation $c_0$ for $t< 0$
and, at time $t=0$, the input parameters change instantaneously to
$\mu_1$, $D_1$, and $c_1$.  At some later time $t_0>0$, the inputs
change back to the original values, $\mu_0$, $D_0$, and $c_0$.  A
small value of $t_0$ models a pulse change in the inputs.  Taking
$t_0=\infty$ models a step change. The discussion here can easily be
generalized to arbitrary time-dependent input (e.g., sinusoidally
varying inputs) by solving a simple linear ODE for the time
dependent mean, variance, and covariance~\citep{gardiner85}.

We assume that for time $t\le0$, the distribution is in its steady state so that
$$
\left.\begin{array}{lcl}m(t)&=&\mu_0,\medskip\\
\sigma^2(t)&=&D_0\medskip\\
\gamma(t)&=&cD_0
\end{array}\;\right\rbrace\quad t\le 0
$$
At time $t=0$, the input statistics change and the mean and covariance matrix begin to track this change.  In particular,
$$
\left.\begin{array}{lcl}m(t)&=&e^{-t}\mu_0+\left(1-e^{-t}\right)\mu_1,\medskip\\
\sigma^2(t)&=&e^{-2t}D_0+\left(1-e^{-2t}\right)D_1\medskip\\
\gamma(t)&=&e^{-2t}c_0D_0+\left(1-e^{-2t}\right)c_1D_1
\end{array}\;\right\rbrace\quad t\in [0,t_0]
$$
At time $t_0$, the input statistics change back to $\mu_0$, $D_0$, and $c_0$ and the distribution relaxes back to its steady state.  In particular,
$$
\left.\begin{array}{cc}m(t)&=e^{-(t-t_0)}m(t_0)+\left(1-e^{-(t-t_0)}\right)\mu_0,\medskip\\
\sigma^2(t)&=e^{-2(t-t_0)}\sigma^2(t_0)+\left(1-e^{-2(t-t_0)}\right)D_0\medskip\\
\gamma(t)&=e^{-2(t-t_0)}\gamma(t_0)+\left(1-e^{-2(t-t_0)}\right)c_0D_0
\end{array}\;\right\rbrace\quad t\ge t_0
$$
where $\mu(t_0)$, $\sigma^2(t_0)$ and $\gamma(t_0)$ are given by the
previous set of equations. The variance and covariance of the
solutions change with a time constant that is twice as fast as the
time constant with which mean changes.  This is a known property of
Ornstein-Uhlenbeck processes~\citep{gardiner85}.

\subsubsection{The time-dependent firing rate}

We now investigate how the firing rate changes in response to a pulse or a step change in the input statistics.  The firing rate at time $t$ is given by $\nu(t)=J(\mu(t),\sigma^2(t),D_t)$ where $J(\mu,\sigma^2,D)$ is defined in \eqref{E:nu}, $\mu(t)$ and $\sigma^2(t)$ are as derived above and $$D_t=\begin{cases}D_0 & t< 0\\ D_1 & t\in[0,t_0]\\ D_0 & t>t_0\end{cases}$$ is the time dependent diffusion coefficient.

We can simplify the expression to get
\begin{align*}
\nu_V(t)&=
 \begin{cases}\frac{\alpha(t)}{\sqrt{\pi}}e^{-\alpha^2(t)} & t< 0\\ \noalign{\medskip}
 \left(\frac{D_1}{e^{-2t}D_0+\left(1-e^{-2t}\right)D1}\right) \frac{\alpha(t)}{\sqrt{\pi}}e^{-\alpha^2(t)} & t\in[0,t_0] \\ \noalign{\medskip}
\left(\frac{D_0}{\left(1+e^{-2t}-e^{-2(t-t_0)}\right)D_0+\left(e^{-2(t-t_0)}-e^{-2t}\right)D_1}\right) \frac{\alpha(t)}{\sqrt{\pi}}e^{-\alpha^2(t)} & t>t_0
\end{cases}
\end{align*}
where
$$
\alpha(t)=\frac{1-m(t)}{\sqrt{2\sigma^2(t)}}.
$$
Note that $\alpha(t)$ changes continuously with $t$.  Thus any discontinuities in the expression above are from the factors multiplying the $\frac{\alpha(t)}{\sqrt{\pi}}e^{-\alpha^2(t)}$ term.  In particular, the firing rate jumps discontinuously by a factor of $\frac{D_1}{D_0}$ at time $0$ and by a factor of $\frac{D_1}{e^{-2t_0}D_0+\left(1-e^{-2t_0}\right)D_1}$ at time $t_0$.
If we change the mean of the input signal, but do not change the variance of the input signal (by setting $\mu_0\ne \mu_1$ and $D_0=D_1$), then the firing rate changes continuously with time constant $\frac{1}{g}=1$.
If, instead, we change $D$ and keep $\mu$ constant (by setting $\mu_0= \mu_1$ and $D_0\ne D_1$), the firing rate has jump discontinuities at time 0 and $t_0$, and changes with a faster time constant of $\frac{1}{2g}=\frac{1}{2}$.

\subsubsection{The time dependent cross-covariance}

We now look at the effects of changes in the input parameters on the conditional firing rate, $\nu_{V|W}(\tau,t)$.
We first derive look at lag $\tau=0$.
The quantity $\nu_{V|W}(0,t)$ quantifies the tendency of the neurons to fire together.
The conditional distribution, $P(t,\V\,|\,\W(t)=1)$, of $\V(t)$ given that $\W$ crossed threshold at time $t$ is a Gaussian with mean and variance given respectively by
$$
m_c(0,t)=m(t)+\rho(t)(1-m(t))
$$
and
$$
\sigma_c^2(0,t)=\sigma^2(t)(1-\rho(t))
$$
where $\rho(t)=\frac{\gamma(t)}{\sigma^2(t)}$ is the sub-threshold correlation and $m(t)$, $\sigma^2(t)$, and $\gamma(t)$ are derived in the previous subsection.
The firing rate at lag $\tau=0$ is then given by
$$
\nu_{V|W}(0,t)=\nu(m_c(0,t),\sigma^2_c(0,t),D_t).
$$

We now derive the conditional firing rate for times $t>0$ and lags $\tau>0$.  We break the derivation into three cases.
The distribution of $V(t+\tau)$ conditioned on a spike in $W$ at time $t$ is a one dimensional Gaussian.
If $t+\tau<t_0$, the mean and variance of this Gaussian are given by
$$
\left. \begin{array}{cc}
m_c(\tau,t)&=e^{-\tau}m_c(0,t)+(1-e^{-\tau})\mu_1\\
\sigma^2_c(\tau,t)&=e^{-2\tau}\sigma_c^2(0,t)+(1-e^{-2\tau})D_1
\end{array}\right\rbrace \,t\in[0,t_0],\; t+\tau\le t_0
$$
If $t\in [t,t_0]$, but $t+\tau>t_0$, the mean and variance are
$$
\left. \begin{array}{cc}
m_c(\tau,t)&=e^{-((t+\tau)-t_0)}m_c(t_0-t,t)+(1-e^{-((t+\tau)-t_0)})\mu_0\\
\sigma^2_c(\tau,t)&=e^{-2((t+\tau)-t_0)}\sigma_c^2(t_0-t,t)+(1-e^{-2((t+\tau)-t_0)})D_0
\end{array}\right\rbrace \,t\in[0,t_0],\; t+\tau>t_0.
$$
Finally, when $t>t_0$, the mean and variance are
$$
\left. \begin{array}{cc}
m_c(\tau,t)&=e^{-\tau}m_c(0,t)+(1-e^{-\tau})\mu_0\\
\sigma^2_c(\tau,t)&=e^{-2\tau}\sigma_c^2(0,t)+(1-e^{-2\tau})D_0
\end{array}\right\rbrace \,t>t_0.
$$
The conditional firing rate is then given by
$$
\nu_{V|W}(\tau,t)=J(m_c(\tau,t),\sigma^2_c(\tau,t),D_{t+\tau}).
$$

\section*{Acknowledgements}

This work was supported by NSF Grants DMS-0604429 and DMS-0817649 and a Texas ARP/ATP award.

\appendix


\end{document}